\documentclass[aps,prl,reprint,superscriptaddress,twocolumn]{revtex4}

\usepackage{graphicx,amsmath,amssymb,dcolumn,physics,color}

\begin{document}

\title{Scaling of Energy Dissipation in Nonequilibrium Reaction Networks}

\author{Qiwei Yu}
\affiliation{School of Physics, Peking University, Beijing 100871, China}

\author{Dongliang Zhang}
\affiliation{School of Physics, Peking University, Beijing 100871, China}

\author{Yuhai Tu}
\affiliation{IBM T.~J.~Watson Research Center, Yorktown Heights, NY 10598}

\date{\today}

\begin{abstract}
The energy dissipation rate in a nonequilibirum reaction system can be determined by the reaction rates in the underlying reaction network. By developing a coarse-graining process in state space and a corresponding renormalization procedure for reaction rates, we find that energy dissipation rate has an inverse power-law dependence on the number of microscopic states in a coarse-grained state. The dissipation scaling law requires self-similarity of the underlying network, and the scaling exponent depends on the network structure and the flux correlation. Implications of this inverse dissipation scaling law for active flow systems such as microtubule-kinesin mixture are discussed.
\end{abstract}

\maketitle

\clearpage

Living systems are far from equilibrium. Energy dissipation is critical not only for growth and synthesis but also for more subtle information processing and regulatory functions. The free energy dissipation is directly related to the violation of detailed balance--a hallmark of nonequilibrium systems--in the underlying biochemical reaction networks~\cite{Gnesotto2018}.
In particular, driven by energy dissipation (e.g., ATP hydrolysis), these biochemical systems can reach nonequilibrium steady states (NESS) that carry out the desired biolgical function. One of the fundamental questions is then how much energy dissipation is needed for performing certain biological function. Indeed, much recent research has been devoted to understanding the relation between the energy cost and the performance of biological functions such as sensing and adaptation~\cite{Lan2012,Mehta2012}, error correction~\cite{Hopfield1974,Bennett1979}, accurate timing in biochemical oscillations~\cite{Cao2015} and synchronization~\cite{Zhang2019}.

Quantitatively, the free energy dissiaption rate can be determined by computing the entropy production rate in the underlying stochastic reaction network given the transition rates between all microscopic states of the system~\cite{hill_1977,Qian2006}. However, for complex systems with a large number of microscopic states, the system may only be measured at a coarse-grained level with coarse-grained states and coarse-grained transition rates among them. By following the same procedure for computing entropy production rate, we can determine the energy dissipation rate at the coarse-grained level. 
An interesting question is whether and how the energy dissipation rate at a coarse-grained level is related to the ``true" energy dissipation rate obtained at the microscopic level. 
Here, we attempt to connect the dissipation at different scales by developing a coarse-graining procedure inspired by the real space renormalization group (RG) approach by Kadanoff~\cite{Kadanoff1966, Wilson1975} and applying it to various reaction networks in the general state space, which can include both physical and chemical state variables.

{\bf Nonequilibrium reaction network.} Each node in the reaction network represents a state of the system and each link represents a reaction with the transition rate from state $i$ to state $j$ given by 
\begin{equation}
	k_{i,j} =  k_{i,j}^0 \gamma_{i,j} =  \frac{2k_0}{1+\exp\qty({\Delta E_{i,j}/k_BT})} \gamma_{i,j}, 
\end{equation}
where $k_{i,j}^0$ represents the equilibrium reaction rates and $\Delta E_{i,j} (= E_i-E_j)$ is the energy difference between states $i$ and $j$. We set $k_0=1$ for the time scale and $k_B T=1$ for the energy scale. The equilibrium rates satisfy detailed balance $k_{i,j}^0/k_{j,i}^0 =e^{-\Delta E_{i,j}}$ and $\gamma_{i,j}$ represents the nonequilibrium driving force. For a given loop $(l_1,l_2,...,l_n,l_1)$ of size $n$ ($l_{n+1}=l_1$) in the network, we define a nonequilibrium parameter $\Gamma$ as the ratio of the product of all the rates in one direction over that in the reverse direction: $\Gamma =  \Pi_{k=1}^{n}\frac{\gamma_{l_{k+1},l_k}}{\gamma_{l_k,l_{k+1}}}$. The system breaks detailed balance if there is one or more loops for which $\Gamma\ne 1$. The steady-state probability distribution $\{P_i^{ss}\}$ can be solved from the master equation:
	$\sum_{j}\qty(k_{j,i} P_{j}^{ss}-k_{i,j} P_{i}^{ss})=0 $ with normalization $\sum_i P_i^{ss}=1$.
The steady-state dissipation (entropy production) rate is given by~\cite{hill_1977,Qian2006}:
\begin{equation}
	\dot{W}=\sum_{i<j} (J_{i,j}-J_{j,i}) \ln\frac{J_{i,j}}{J_{j,i}},
	\label{Equ: total dissipation}
\end{equation}
where $J_{i,j}=k_{i,j} P^{ss}_{i}$ is the steady-state probability flux from state $i$ to state $j$. Here, we consider the simplest case with a flat energy landscape ($\Delta E_{i,j}=0$) and a random nonequilibrium force $\gamma_{i,j}$ that follows a lognormal distribution: $\ln \left(\gamma_{i,j}\right)\sim\mathcal{N}\left(\mu, \sigma\right)$ (other distributions are used without affecting the results, see SI for details). 

\begin{figure}
	\centering
	\includegraphics[width=\linewidth]{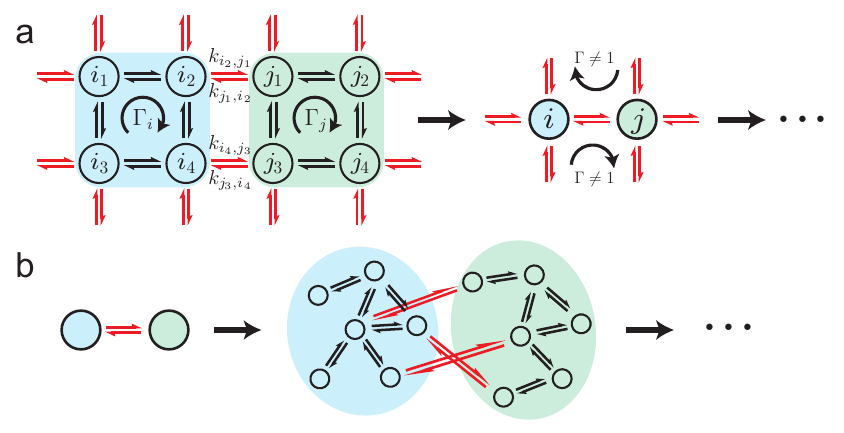}
	\caption{
		(a) Illustration of the coarse-graining process in square lattice. All states in the shaded area (blue or green) are merged to form the new CG state. The red links are combined together to form the transition reaction between the new states, while black links correspond to internal transitions that are removed in the CG model.
		(b) Illustration of the growth mechanism in random hierarchical network. The example here corresponds to $m=6$, $d=2$. 
	}
	\label{Fig: schematics of the model}
\end{figure}

\textbf{State space renormalization and energy dissipation scaling.} The network can be coarse-grained by grouping subsets of highly connected (neighboring) states to form a coarse-grained (CG) state while conserving both total probability of the state and the total probability flux between states. 
For example, when we group two sets of microscopic states, $(i_1,i_2,$\textellipsis$ ,i_r)$ and $(j_1,j_2,$\textellipsis$ ,j_r)$, to form two CG states $i$ and $j$, the probability of each CG state is the sum of the probability of all the constituent states: 
\begin{equation}
	P^{ss}_i = \sum_{\alpha=1}^{r} P^{ss}_{i_\alpha}, \ \ P^{ss}_j = \sum_{\alpha=1}^{r} P^{ss}_{j_\alpha}.
	\label{Pss}
\end{equation}
The transition rates in the CG system is renormalized to preserve the total probability flux from state $i$ to $j$: 
\begin{equation}
	k_{i,j} = \frac{J_{i,j}}{P^{ss}_i} = \frac{1}{{P^{ss}_i}}\sum_{(\alpha, \beta)} J_{i_\alpha, j_\beta} = \frac{\sum_{(\alpha, \beta)} k_{i_\alpha, j_\beta}P^{ss}_{i_\alpha}}{\sum_{\alpha=1}^{r} P^{ss}_{i_\alpha}}.
	\label{k_ij}
\end{equation}
Fig.~\ref{Fig: schematics of the model}a demonstrates an example in a square lattice with $r=4$. 
The red links correspond to transitions that survive the coarse-graining process with their reaction rates renormalized according to Eq.~\ref{k_ij}. The black links represent internal transitions that are averaged over during coarse-graining. The dissipation rate of the CG system can be computed from Eq.~\ref{Equ: total dissipation} with the renormalized probability distribution (Eq.~\ref{Pss}) and transition rates (Eq.~\ref{k_ij}).  

For a microscopic system with $n_0$ states, coarse-graining $s$ times leads to a system with $n_s$ states. Each state in the CG system hence contains $\frac{n_0}{n_s}$ original states. We define  
$\frac{n_0}{n_s}$ as the block size, which is used to characterize the degree (scale) of coarse-graining. Our main result is that the dissipation rate of the CG system $\dot{W}(n_s)$ 
scales as an inverse power law with respect to the block size for a diverse class of reaction networks: 
\begin{equation}
	\frac{\dot{W}(n_s)}{ \dot{W}(n_0)} = \left(\frac{n_0}{n_s}\right)^{-\lambda} ,
	\label{Equ: definition of scaling law}
\end{equation}
where $\lambda$ is the dissipation scaling exponent.
Furthermore, the exponent $\lambda$ depends on the structure of the network with an unifying expression for the networks we studied:
\begin{equation}
	\lambda = d_{L} - \log_r (1+C^*),
	\label{Equ: lambda general case}
\end{equation}
where $r=n_{s}/n_{s+1}$ is the number of fine-grained states in a next-level CG state and the link exponent $d_{L}$ is defined as the scaling exponent of the total number of links (reactions) $L$ with respect to the block size:
\begin{equation}
	d_L  \equiv \frac{\ln(L(n_s)/L(n_0))}{\ln (n_s/n_0)}.
	\label{Equ: deff definition}
\end{equation}
The detailed derivation of Eq.~\ref{Equ: lambda general case} is provided in the SI. The parameter $C^*$ is the average correlation between probability fluxes given by:
\begin{equation}
	C^* = \frac{\langle A_{i_\alpha,j_\beta} \qty(\ A_{i,j} - A_{i_\alpha,j_\beta})\rangle_{i_\alpha,j_\beta}}{\sqrt{\langle A_{i_\alpha,j_\beta}^2\rangle_{i_\alpha,j_\beta} \langle \qty(\ A_{i,j} - A_{i_\alpha,j_\beta})^2 \rangle_{i_\alpha,j_\beta} }},
	\label{C*}
\end{equation}
where $A_{x,y} = J_{x,y}-J_{y,x}$ is the net probability flux between states $x$ and $y$, and $\qty(A_{i,j} - A_{i_\alpha,j_\beta})$ is the sum of all other fluxes that are merged with $ A_{i_\alpha,j_\beta}$ during coarse-graining. Next, we demonstrate the energy dissipation scaling in three different types of extended networks.

\begin{figure}
	\centering
	\includegraphics[width=\linewidth]{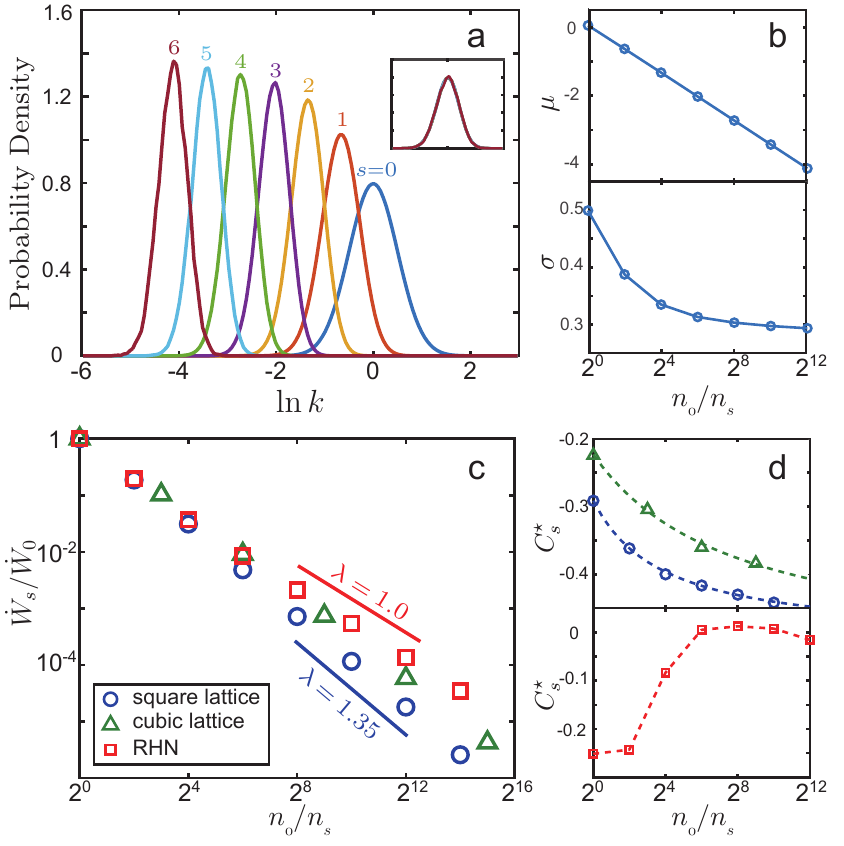}
	\caption{
		(a)Probability density function (PDF) of $\ln k_{i,j}$ at different CG levels (from left to right, coarse-grained to fine-grained). Inset: normalized PDF all collapse to a standard Gaussian distribution.
		(b) Mean ($\mu$) and standard deviation ($\sigma$) of the $\ln k_{i,j}$ distribution as a function of the block size $n_0/n_s$.
		(c) Power-law relation between the scaled dissipation rate $\dot{W}_s/\dot{W}_0$ and the block size $n_0/n_s$, in square lattice (blue circle), cubic lattice (green triangle), and random hierarchical network (red square, $d=m=4$).  
		(d) Correlation coefficient $C^*_s$ of the three systems plotted in (c).
	}
	\label{Fig: regular and RHN results}
\end{figure}

\textbf{Regular lattice.} We start our analysis with a $N_0\times N_0$ square lattice where transitions can only take place between nearest neighbors. The coarse-graining is carried out by grouping $4 (=2\times 2)$ neighboring states at one level to create a CG state at the next level iteratively (Fig.~\ref{Fig: schematics of the model}a). Periodic boundary conditions are imposed to prevent any boundary effects (see Fig.~S1a).

Both transition rates and the overall dissipation evolve as the system is coarse-grained. As shown in Fig.~\ref{Fig: regular and RHN results}a, the renormalized transition rates follow lognormal distributions at all CG levels, i.e. $\ln{k}\sim\mathcal{N}(\mu,\sigma)$, with mean and variance decreasing with the block size (Fig.~\ref{Fig: regular and RHN results}b). Interestingly, the mean $\mu$ decreases by $\ln{2}$ after each coarse-graining, effectively doubling the timescale after the length scale is doubled, which indicates that transitions between CG states are slower. Consequently, it is expected that the dissipation rate also decreases with coarse graining. Remarkably, the dissipation rate decreases with the block size by following a power-law (Fig.~\ref{Fig: regular and RHN results}c, blue circle). The numerically determined scaling exponent $\lambda_\mathrm{2d}=1.35$ suggests that dissipation actually decreases faster than the block size, which can be rationalized with Eq.~\ref{Equ: lambda general case}. Since the block size at the $s$-th level is $4^s$, we have $r=4$. The number of links is inversely proportional to the block size, giving the link exponent $d_{L}=1$. The resultant scaling exponent is
\begin{equation}
	\lambda_\mathrm{2d} = 1- \log_4\qty(1+C^*),
	\label{Equ: square lattice lambda}
\end{equation}
where $C^*$ denotes the Pearson correlation coefficient of the probability fluxes defined in Eq.~\ref{C*}.  
For the example shown in Fig.~\ref{Fig: schematics of the model}a, it is given by 
\begin{equation}
	C^* = \frac{\langle \qty(J_{i_2,j_1} -J_{j_1,i_2})\qty(J_{i_4,j_3} -J_{j_3, i_4})\rangle}{\sqrt{\langle \qty(J_{i_2,j_1} -J_{j_1,i_2})^2 \rangle \langle \qty(J_{i_4,j_3} -J_{j_3, i_4})^2\rangle}},
\end{equation}
where the correlation is averaged over all such pairs $(i,j)$ in the entire lattice.  $C^*$ can be directly calculated from the fluxes (Fig.~\ref{Fig: regular and RHN results}d, blue circles). It appears to decrease with the block size and converge to a fixed point  $\sim -0.50$ (by extrapolation), which corresponds to a scaling exponent of $\lambda=1.50$ at the infinite size limit. 
For the finite systems studied here, the correlation coefficient $C^*$ is  larger than its infinite size value, and 
the exponent found in our simulations is slightly smaller 
($\lambda_\mathrm{2d}=1.35<1.50$).

The 2D results above can be generalized to regular lattice in higher dimensions, where $d_{L}$ remains $1$ and the correlation coefficient $C^*$ converges to a fixed point analogous to the 2D case. 
For example, the numerically determined scaling exponent in the cubic regular lattice is  $\lambda_\mathrm{3d}=1.23>1$  (Fig.~\ref{Fig: regular and RHN results}c, green triangles). The correlation coefficient $C^*$ in 3D is found to converge to a value slightly greater than its 2D value (Fig.~\ref{Fig: regular and RHN results}d, green triangles). Therefore, the scaling exponent in the cubic lattice is slightly smaller than that in the square lattice. 

Overall, the dissipation rate decays with the block size with an exponent $\lambda$ that is larger than the link exponent of the network for regular lattice network due to the negative probability flux correlation  $C^*$, which is caused by the highly regular structure of the lattice network.  For random reaction networks, this correlation vanishes as evidenced by the case discussed next.

\textbf{Random hierarchical network.} To investigate the dissipation scaling behavior in networks with irregular but self-similar structures, we introduce a  generalization of the regular lattice called random hierarchical network (RHN). It shares many features of the regular lattice, such as the conservation of average degree at different CG levels. However, the links among neighboring states in RHN are created randomly to disrupt the local regularity of the network. 
Specifically, RHN is constructed from a small initial network with an iterative growth mechanism (see Fig.~\ref{Fig: schematics of the model}b).
We start at the coarsest level with an initial network that has $n_s$ states and $(n_sd)/{2}$ random links (average $d$ links per state), and grow it for $s$ times to obtain the finest level network with $n_0$ states.
In each growth step, each macro-state splits into $m$ micro-states with $\qty(md)/{4}$ links randomly created among them. Each link then splits into $m/2$ links by randomly choosing $m/2$ distinct pairs of micro-states that belong to the two macro-states and connecting them pairwise. In this way, the average degree $d$ is preserved in all of the CG levels. Each growth step results in an $m$-fold increase in both the number of states and the number of links, leading to $r=m$ and $d_{L}=1$. After reaching the finest level, we assign the transition rates according to the lognormal distribution as before, determine the steady-state probability distribution, and coarse-grain the system by precisely reversing the growth procedure. 

The dissipation rate in RHN also scales with the block size in a power-law manner (Fig.~\ref{Fig: regular and RHN results}c, red squares) with  
the scaling exponent $\lambda_\mathrm{RHN}\approx 1$ regardless of the choices of parameters used to specify the growth procedure, namely $d$ and $m$ (see Table S1 in SI for details). In other words, the dissipation always scales the same as the number of states. In RHN, the net flux correlation $C^*$ vanishes at the RG fixed point (Fig.~\ref{Fig: regular and RHN results}d, red squares) due to the randomness of the reaction links. Therefore, according to the general expression for the scaling exponent (Eq.~\ref{Equ: lambda general case}), we have $\lambda_\mathrm{RHN}=d_{L}=1$ independent of $d$ or $m$.


The RHN can be considered as a mean-field generalization of a regular lattice of dimension $\log_2m$. In both cases, each coarse graining operation leads to a $m$-fold decrease in the number of states as well as the total number of links. Therefore, both the regular lattice and RHN have the same link exponent $d_{L}=1$. Their different dissipation scaling exponents comes from the different values of net flux correlation $C^*$.   Next, we study how the dissipation scaling depends on the topology of the network characterized by the link exponent $d_{L}$. 

\begin{figure}
	\centering
	\includegraphics[width=\linewidth]{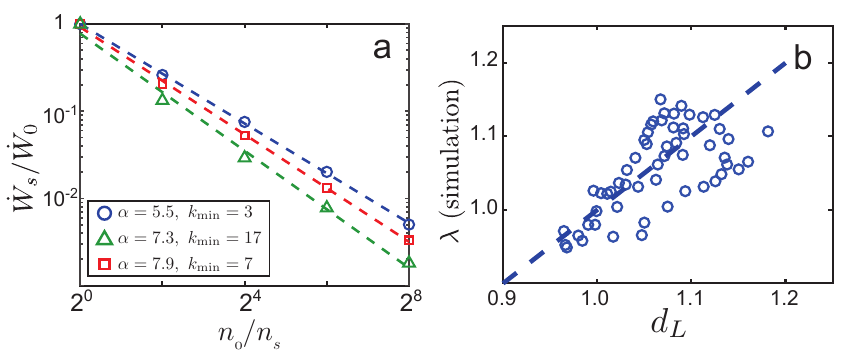}
	\caption{
		(a) The scaled dissipation rate $\dot{W}_s/\dot{W}_0$ in three different scale-free networks. 
		(b) The dissipation scaling exponent in simulation is positively correlated with the link exponent $d_{L}$, in $56$ different scale-free networks. The dashed line corresponds to perfect agreement $\lambda=d_{L}$. The correlation coefficient is $0.65$.
	}
	\label{Fig: scale-free results}
\end{figure}

\textbf{Scale-free network.}
We consider scale-free networks (SFN) characterized by a power-law degree distribution $p(k)\propto k^{-\alpha}$ ($k\geq k_\mathrm{min}$) \cite{Barabsi1999, Barabsi2009}. To elucidate the dissipation scaling behavior in comparison with the networks studies above, we embed the  SFN in a 2D plane and conduct the coarse-graining process as in the square lattice~\cite{Rozenfeld2002,Kim2004}. Briefly, the network is constructed by assigning degrees to all sites on a square lattice according to a power-law distribution and fulfilling the degree requirements by considering the sites in a random order. For each site, we examine its neighbors from close to distant and create links whenever possible, until the number of its links reaches its pre-assigned degree or the search radius reaches an upper limit. The preference to connecting with nearer neighbors allows us to use the coarse-graining method that group neighboring states together. Although not all degree requirements can be satisfied, the resulted network is indeed scale-free, consistent with previous work~\cite{Kim2004}.

The dissipation rate in the 2D-embedded SFN also scales with block size as a power law with the exponent $\lambda$ depending on the network structure (Fig.~\ref{Fig: scale-free results}a). This can be explained by considering the two terms contributing to the scaling exponent, i. e., the link exponent $d_{L}$ and flux correlation coefficient $C^*$. Due to the local randomness in SFN, we expect $C^*\approx 0$ as in RHN. The energy dissipation scaling exponent is then determined by $d_{L}$, which can be determined by the fractal dimension $d_B$ and power exponent $\alpha$ of the embedded SFN (see SI for detailed derivation): 
\begin{equation}
	\lambda_\mathrm{SFN}\approx d_{L} \approx 1+\frac{d_B-2}{2\qty(\alpha-1)}. 
	\label{Equ: deff in SF}
\end{equation}
The fractal dimension $d_B$ depends on the degree exponent $\alpha$ and the minimum degree $k_\mathrm{min}$~\footnote{Previous work has shown that the CG system converges to a (trivial) complete-graph fixed point when $\alpha$ is small and to a fractal-network fixed point when $\alpha$ is large~\cite{Rozenfeld2010}. Here we choose sufficiently large $\alpha$ to ensure that the CG system is fractal and nontrivial. }. It can be numerically determined by calculating the minimum number of boxes of size $l_B$ needed to cover the entire network at the finest level \cite{Song2007, Song2005,Kim2007}.

To test the estimation of $\lambda_\mathrm{SF}$, we constructed 56 scale-free networks with $\alpha\in[5,7]$ and $k_\mathrm{min}\in[3,22]$ and computed the energy dissipation scaling exponent. As shown in Fig.~\ref{Fig: scale-free results}b, there is a positive correlation between $d_{L}$ and $\lambda_\mathrm{SF}$. The deviations from the diagonal are likely caused by residual correlation $C^*$ that has not completely vanished during the limited number of coarse-graining operations in our simulations. 
The regular 2D-embedding may also create some initial correlations. 

{\bf Scaling requires network self-similarity.} The dissipation scaling does not exist in all networks. For example, even though the dissipation rate decreases with coarse-graining in both Watts-Strogatz small-world network~\cite{Watts1998} and the Erd{\H{o}}s-R{\'{e}}nyi random network~\cite{Albert2002}, the scaling law defined by Eq.~\ref{Equ: definition of scaling law} is not satisfied in either of these networks (see Fig.~S8 in SI for details). 
The existence of the dissipation scaling law depends on whether the network has self-similarity, i.e., whether the CG process converges the network to the complete-graph fixed point or a self-similar (fractal) fixed point~\cite{Rozenfeld2010}. The regular lattices, RHN, and SFN converge to a self-similar fixed point, i.e.,  networks at all CG levels are structurally similar and properties like the number of links (reactions) and total dissipation rate all scale in a power-law fashion. However, in the small-world network or the Erd{\H{o}}s-R{\'{e}}nyi network, the CG process eventually generates a complete graph with all nodes directly connected. The resultant complete graph bears no resemblance to the original network structure, and the scaling properties are thus absent. For self-similar networks, the scaling exponent $\lambda$ depends on the link exponent $d_{L}$ and the flux correlation coefficient $C^*$. While $d_{L}$ reflects the global self-similarity across different levels, $C^*$ quantifies the correlation between parallel fluxes which is nonzero only when there is certain regularity in the local links.

\begin{figure}
	\centering
	\includegraphics[width=\linewidth]{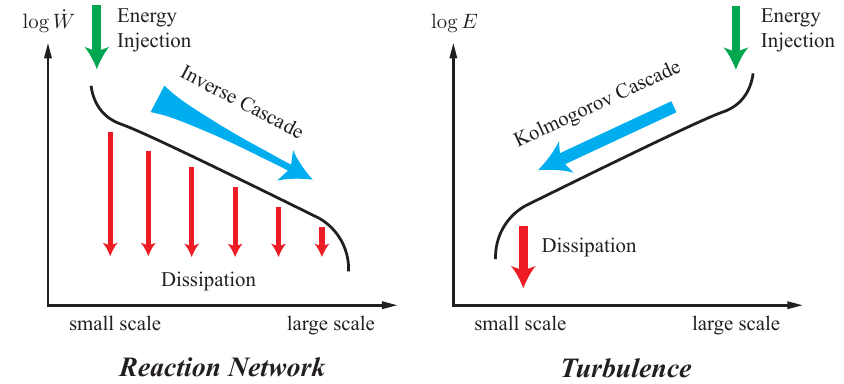}
	\caption{
		Comparison between the inverse energy dissipation cascade in self-similar reaction networks and the Kolmogorov energy cascade in turbulence. See text for detailed discussion. 
	}
	\label{Fig:illu}
\end{figure}

{\bf Discussion.} The microscopic state variable considered here is general and can include both chemical state of a molecule such as its phosphorylation state that can be changed by energy consuming enzymes (kinase and phosphatase) as well as its physical location  that can be transported by molecular motors that consume ATP. 
The dissipation scaling in self-similar reaction networks is reminiscent of the Kolmogorov scaling theory in homogeneous turbulence, which is also based on self-similarity of the turbulence structures (``eddies") at different scales in the inertia range~\cite{Pope2000}. 
However, as illustrated in Fig.~\ref{Fig:illu}, there are fundamenal differences -- one is about scaling of the energy spectrum in turbulence and the other is on scaling of energy dissipation rate in nonequilibrium reaction networks. Specifically, while energy is introduced at large length scale in turbulence, free energy is injected at the microscopic scale in reaction networks, which leads to the ``inverse cascade" of energy dissipation from small scale to large scale. Second, while energy is conserved within the inertia range in turbulence, it is dissipated at all scales in nonequilibrium networks. In fact, the inverse scaling law, Eq.~\ref{Equ: definition of scaling law}, indicates that the energy dissipation rate in a coarse-grained network (CGN) is much lower than that in its preceding fine-grained network (FGN). The difference in energy dissipation in CGN and FGN is due to two ``hidden" free energy costs in CGN: 1) the energy dissipation needed to maintain the NESS of a CG state, which contains many internal microscopic states and transitions among them; 2) the entropy production due to merging multiple reaction pathways into a CG transition (reaction) between two CG states in the coarse-graining process~\cite{Santillan2011,Esposito2012} (See SI section III for details of the energy dissipation partition).

Accroding to our results here, the energy dissipation of a nonequilibrium system determined from its dynamics at a CG scale can be significantly smaller than the true energy cost at the microscopic scale. In the active microtubule-kinesin system, for example, ATP is hydrolyzed to drive the relative motion of microtubules (MT) with the microscopic coherent length given by the kinesin persistent run length $l_0\sim 0.6-1\mathrm{\mu m}$~\cite{Vale1996Direct,Verbrugge2009Novel}. The active flow of the MT-kinesin system can occur at a much larger length scale $l_{f} \sim 100\mu\mathrm{m}$~\cite{Sanchez2012}. 
Therefore, the energy dissipation rate $\dot{W}_f$ determined (estimated) at the length scale of the active flow $l_{f}$ can be related to the true energy dissipation rate $\dot{W}_0$ at the microscopic scale $l_0$ by using the energy dissipation scaling law (Eq.~\ref{Equ: definition of scaling law}):  $\frac{\dot{W}_f}{\dot{W}_0}\approx ((l_0/l_{f})^3)^{\lambda_{3d}}\approx 10^{-7.4}-10^{-8.2}$, which means that 
most of the energy is spent to generate and maintain the flow motion at different length scales from $l_0$ to $l_f$, and only a tiny amount is used to overcome viscosity (of the medium) at the large flow scale $l_{f}$. 
Realistic active systems contain microscopic details not included in the simple models studied here, e.g., the transition (transport) rates are determined by dynamics of motor molecules, MT has its specific spatial structure, and the energy landscape may not be flat ($E_i\ne E_j$), etc. 
Whether and how these realistic microscopic interactions change the dissipation energy scaling remain interesting open questions. 

{\bf Acknowledgments.} Y.~T. acknowledges stimulating discussions with Drs. Dan Needleman and Peter Foster, whose measurements of heat dissipation in active flow systems partly inspired this work. The work by Y.~T. is supported by a NIH grant (5R35GM131734 to Y.~T.). 
Q.~Y. thanks the hospitality of Center for Theoretical Biological Physics, Rice University. 

\bibliography{scaling_citations}

\end{document}


\title{Supporting Information: Scaling of Energy Dissipation in Nonequilibrium Reaction Networks} 
\author{Qiwei Yu}
\affiliation{School of Physics, Peking University, Beijing 100871, China}

\author{Dongliang Zhang}
\affiliation{School of Physics, Peking University, Beijing 100871, China}

\author{Yuhai Tu}
\affiliation{Physical Sciences Department, IBM T.~J.~Watson Research Center, Yorktown Heights, NY 10598}

\date{\today}

\maketitle
\tableofcontents
\newpage
\section{Supplementary analytic derivation}
\subsection{Derivation of the scaling exponent $\lambda$}
To derive the expression for the scaling exponent $\lambda$ in the main text, we will first study the square lattice, where the derivation is more intuitive, and then generalize it to generic reaction networks.

\subsubsection{The square lattice}
The steady-state dissipation rate of the system is given by
\begin{equation}
	\dot{W}=\sum_{i<j}\left(J_{i,j}-J_{j,i}\right) \ln \frac{J_{i,j}}{J_{j,i}}=\sum_{i<j}\left(k_{i,j} P_{i}^{ss}-k_{j,i} P_{j}^{ss}\right) \ln \frac{k_{i,j} P_{i}^{ss}}{k_{j,i} P_{j}^{ss}},
	\label{Equ: total dissipation}
\end{equation}
where $P_i^{ss}$ is the stead-state probability of state $i$ and $J_{i,j}$ is the steady-state flux from state $i$ to state $j$~\cite{hill_1977}. The transition fluxes can be decomposed into symmetric and antisymmetric components:
\begin{equation}
	J_{i,j} = \frac{1}{2} \qty(S_{i,j} + A_{i,j}),
\end{equation}
where $S_{i,j}=J_{i,j}+J_{j,i}$ and $A_{i,j} = J_{i,j}-J_{j,i}$. The antisymmetric component $A_{i,j}$ is actually the net current from state $i$ to $j$. The dissipation rate as function of $A$ and $S$ reads
\begin{equation}
	\dot{W}=\sum_{i<j} A_{i,j} \ln \frac{S_{i,j}+A_{i,j}}{S_{i,j}-A_{i,j}}
\end{equation}
At the microscopic scale, we take the continuum limit that the net flux is an infinitesimal flux compared to the symmetric flux, i.e. $\qty|A_{i,j}|\ll\qty|S_{i,j}| $. This leads to 
\begin{equation}
	\dot{W}=\sum_{i<j} A_{i,j} \ln \frac{S_{i,j}+A_{i,j}}{S_{i,j}-A_{i,j}}
	= \sum_{i<j} A_{i,j} \ln \frac{1+\frac{A_{i,j}}{S_{i,j}}}{1-\frac{A_{i,j}}{S_{i,j}}}
	\approx  2 \sum_{i<j} \frac{A_{i,j}^2}{S_{i,j}} = 2L \langle  \frac{A_{i,j}^2}{S_{i,j}} \rangle,
	\label{Equ: total dissipation in A and S}
\end{equation}
where $L$ is the number of links and $\langle\cdot\rangle$ denotes averaging over all links. As the system is coarse-grained, this approximation is valid as long as the system is not far from equilibrium, which is the case for the flat energy landscape that we study here. Numeric justifications for this approximation will be provided in the later sections of SI. Arguably, higher order terms of $\frac{A}{S}$ must be taken into consideration if the energy landscape is not flat and the fluxes are strongly driven in one particular direction.

As demonstrated in Fig.~1a of the main text, renormalization in the square lattice involves merging two adjacent fluxes:
\begin{equation}
	J_{i,j} = J_{i_2,j_1} + J_{i_4,j_3}.
\end{equation} 
Hence, their symmetric and antisymmetric fluxes are merged accordingly:
\begin{equation}
	\begin{aligned}
	S_{i,j} = S_{i_2,j_1} + S_{i_4,j_3}, \enspace
	A_{i,j} = A_{i_2,j_1} + A_{i_4,j_3}.
	\end{aligned}
\end{equation}
This leads to 
\begin{equation}
	\begin{aligned}
	&\langle S_{i,j} \rangle=  \langle S_{i_2,j_1} \rangle+ \langle S_{i_4,j_3}\rangle = 2\langle S_{i_\alpha,j_\beta} \rangle,\\
	&\langle A_{i,j}^2\rangle = \langle A^2_{i_2,j_1}\rangle +\langle A^2_{i_4,j_3}\rangle +2\langle A_{i_2,j_1}A_{i_4,j_3}\rangle=2\langle A^2_{i_\alpha,j_\beta}\rangle(1+C)
	\end{aligned}
\end{equation}
where $C = \mathrm{corr} (A_{i_2,j_1}, A_{i_4,j_3})$ is the Pearson correlation coefficient between the adjacent fluxes, explicitly given by
\begin{equation}
C^* = \frac{\langle A_{i_2,j_1}A_{i_4,j_3}\rangle}{\sqrt{ \langle A_{i_2,j_1}^2\rangle\langle A_{i_4,j_3}^2 \rangle}} = \frac{\langle \qty(J_{i_2,j_1} -J_{j_1,i_2})\qty(J_{i_4,j_3} -J_{j_3, i_4})\rangle}{\sqrt{\langle \qty(J_{i_2,j_1} -J_{j_1,i_2})^2 \rangle \langle \qty(J_{i_4,j_3} -J_{j_3, i_4})^2\rangle}},
\end{equation}
Quantities with subscripts $i_\alpha$ and $j_\beta$ are averaged over microscopic links while those with subscripts $i$ and $j$ are averaged over macroscopic (coarse-grained) links. The link-averaged term $\langle A^2/S\rangle$ in Eq.~\ref{Equ: total dissipation in A and S} is thus given by
\begin{equation}
	\frac{\left\langle  \frac{A_{i,j}^2}{S_{i,j}} \right\rangle}{\left\langle  \frac{A_{i_\alpha,j_\beta}^2}{S_{i_\alpha,j_\beta}} \right\rangle} \approx \frac{\langle {A_{i,j}^2} \rangle}{\langle  {A_{i_\alpha,j_\beta}^2} \rangle} \frac{\langle S_{i_\alpha,j_\beta}\rangle}{\langle S_{i,j}\rangle} = 1+C.
	\label{Equ: A2 S approximation in the main derivation}
\end{equation}
Note that taking the ratio out of the averaging $\langle\cdot\rangle$ is justified by assuming that the correlation between $A^2$ and $S^{-1}$ does not change in different coarse-grained layers. Even if the correlation between $A^2$ and $S^{-1}$ does change after renormalization, its cumulative contribution to the dissipation is bounded, so the average effect per round goes to zero in the limit of infinite rounds of renormalization (which leads to the RG fixed point).
We will provide further numeric justifications for this approximation in the later sections of the SI. The ratio of dissipation rate in the coarse-grained system to the original one is 
\begin{equation}
	\frac{\dot{W}_1}{\dot{W}_0} = \frac{L_1}{L_0}(1+C_0).
\end{equation} 
Here we add subscript $0$ to $C$ to indicate that it is the correlation at the finest-grained level. The dissipation rate at the $s$-th coarse-grained level is given by 
\begin{equation}
	\frac{\dot{W}_s}{\dot{W}_0} = \frac{L_s}{L_0} \prod_{i=0}^{s-1}(1+C_i) \to \frac{L_s}{L_0} (1+C^\star)^s,
\end{equation} 
where $C^\star$ is the RG fixed point of $C$:
\begin{equation}
	C^\star = \lim\limits_{s\to\infty}C_s
\end{equation}
The block size and the number of links in the system after $s$ rounds of coarse-graining is given by:
\begin{equation}
	\frac{n_0}{n_s} = 4^s, \enspace \frac{L_s}{L_0} = \frac{1}{4^s}
\end{equation}
Therefore, the dissipation scaling exponent is 
\begin{equation}
	\lambda_\mathrm{2d} = -\lim\limits_{s\to\infty} \frac{\ln \frac{\dot{W}_s}{\dot{W}_0}}{\ln \frac{n_0}{n_s}}= \lim\limits_{s\to\infty} \frac{s\ln4-\sum_{i=1}^{s}\ln (1+C_i)}{s\ln4} = 1- \log_4\qty(1+C^\star),
\end{equation}
thus deriving Eq.~9 of the main text.

\subsubsection{The general expression}
In cubic lattice or even more general reaction networks, coarse-graining would involve the merging of more than two fluxes. Suppose the renormalized flux $J_{i,j}$ comprises $t$ microscopic fluxes denoted by $J_{i_1,j_1}$, $J_{i_2,j_2}$, \textellipsis, $J_{i_t,j_t}$:
\begin{equation}
	J_{i,j} = \sum_{\alpha=1}^t J_{i_\alpha,j_\alpha}.
\end{equation} 
Their symmetric and antisymmetric fluxes are merged accordingly:
\begin{equation}
\begin{aligned}
S_{i,j} = \sum_{\alpha=1}^t S_{i_\alpha,j_\alpha}, \enspace
A_{i,j} = \sum_{\alpha=1}^t A_{i_\alpha,j_\alpha}.
\end{aligned}
\end{equation}
We assume that all microscopic fluxes $J_{i_\alpha,j_\alpha}$ follow the same statistics. The summation over $\alpha$ can thus be rewritten as the averaging over the distribution of microscopic fluxes $\langle\cdot\rangle$. This leads to 
\begin{equation}
\begin{aligned}
&\langle S_{i,j} \rangle= \left\langle \sum_{\alpha=1}^t  S_{i_\alpha,j_\alpha} \right\rangle =t\langle S_{i_\alpha,j_\alpha} \rangle,\\
&\langle A_{i,j}^2\rangle =\left\langle \sum_{\alpha=1}^t A^2_{i_\alpha,j_\alpha} + 2 \sum_{\alpha=1}^t\sum_{\beta=\alpha+1}^tA_{i_\alpha,j_\alpha}A_{i_\beta,j_\beta}\right\rangle=t\langle A^2_{i_\alpha,j_\alpha}\rangle(1+C)
\end{aligned}
\end{equation}
where $C = \mathrm{corr} (A_{i_\alpha,j_\alpha}, A_{i,j}-A_{i_\alpha,j_\alpha})$ is the Pearson correlation coefficient between the microscopic net flux $A_{i_\alpha,j_\alpha}$ and the sum of all the microscopic net fluxes that will be merged with it to form the macroscopic flux $A_{i,j}$. Following the same argument used in the square lattice, the link-averaged term $\langle A^2/S\rangle$ is thus multiplied by $(1+C)$ after coarse-graining. The dissipation rate at the $s$-th level is given by 
\begin{equation}
	\frac{\dot{W}_s}{\dot{W}_0} = \frac{L_s}{L_0} \prod_{i=1}^{s}(1+C_i) \to \frac{L_s}{L_0} (1+C^\star)^s,
\end{equation} 
which has a form identical to the square lattice but involves a more general definition of the correlation $C^\star$. The block size and the number of links in the system after $s$ rounds of coarse-graining are given by:
\begin{equation}
\frac{n_0}{n_s} = r^s, \enspace \frac{L_s}{L_0} = \qty(\frac{n_0}{n_s})^{-d_L} = r^{-sd_L}
\end{equation}
where $d_L$ is the link exponent. Therefore, the dissipation scaling exponent is given by 
\begin{equation}
	\lambda =  -\lim\limits_{s\to\infty} \frac{\ln \frac{\dot{W}_s}{\dot{W}_0}}{\ln \frac{n_0}{n_s}}= \lim\limits_{s\to\infty} \frac{sd_L\ln{r}-\sum_{i=1}^{s}\ln (1+C_i)}{s\ln{r}} = d_L- \log_r\qty(1+C^\star),
\end{equation}
recovering Eq.~6 of the main text. The square lattice is a special case with $r=4$ and $d_L=1$. 

\subsection{Derivation of the link exponent $d_L$ in the 2d-embedded scale-free network}
In this part, we estimate the link exponent in the 2d-embedded scale-free network by calculating the number of links in the coarse-grained network. The degree distribution of the initial network is 
\begin{equation}
	p(k) = A k^{-\alpha}, \enspace k\in[k_\mathrm{min},k_\mathrm{max}],
\end{equation}
where $k_\mathrm{min}$ is the minimum degree and $k_\mathrm{max}$ is a sufficiently large cutoff. The normalization constant $A$ is given by
\begin{equation}
	A = \qty(\sum_{k=k_\mathrm{min}}^{k_\mathrm{max}} k^{-\alpha})^{-1} \approx \qty(\int_{k=k_\mathrm{min}}^{+\infty} k^{-\alpha}\dd{k})^{-1}  = \qty(\alpha-1)k_\mathrm{min}^{\alpha-1}.
\end{equation}
The embedding in 2d allows us to use the same coarse-graining method employed in the square lattice. Suppose that we coarse-grain states $i_1$, $i_2$, $i_3$, and $i_4$ to form a new state $i$. The self-similarity property of the scale-free network leads to the following degree relation 
\begin{equation}
	k_{i}= l_B^{-d_k} \max _{\alpha=1, 2, 3, 4} k_{i_\alpha},
\end{equation}
where $l_B=2$ is the linear size of the coarse-grained state and $d_k$ is the degree-scaling fractal dimension of the network~\cite{Song2005}. Since $k_{i_\alpha}$ follows a power-law distribution with minimum degree $k_\mathrm{min}$ and exponent $\alpha$, its cumulative degree distribution is 
\begin{equation}
	P\qty(k_{i_\alpha}\geq k_0) = A \sum_{k=k_0}^{k_\mathrm{max}} k^{-\alpha}\approx A \int_{k=k_0}^{+\infty} k^{-\alpha} \dd{k} =\qty(\frac{k_\mathrm{min}}{k_0})^{\alpha-1}.
	\label{Equ: scale-free CDF}
\end{equation}
The cumulative degree distribution of $k_i$ is therefore:
\begin{equation}
	\begin{aligned}
		P\qty(k_{i}\geq k_0)  &= P\qty(\max _{\alpha=1, 2, 3, 4}   k_{i_\alpha} \geq 2^{d_k}k_0) \\
		&= 1 - \prod_{\alpha=1}^{4}P\qty(k_{i_\alpha}< 2^{d_k}k_0)\\
		&= 1 - \prod_{\alpha=1}^{4}\qty[1- \qty(\frac{k_\mathrm{min}}{2^{d_k}k_0})^{\alpha-1}] \\
		&\approx 4\qty(\frac{k_\mathrm{min}}{2^{d_k}k_0})^{\alpha-1} = \qty(\frac{k'_\mathrm{min}}{k_0})^{\alpha-1}.
	\end{aligned}
	\label{Equ: scale-free renormalized CDF}
\end{equation}
Hence, the coarse-grained network is also scale-free, with the same exponent $\alpha$ but a different minimum degree
\begin{equation}
	k'_\mathrm{min} =  k_\mathrm{min} 2^{\frac{2}{\alpha-1}-d_k}.
\end{equation}
The average degree of nodes in the scale-free network is 
\begin{equation}
	\langle k \rangle =  \sum_{k=k_\mathrm{min}}^{k_\mathrm{max}} p(k)k \approx A \int_{k=k_\mathrm{min}}^{+\infty} k^{1-\alpha}\dd{k} =  \frac{\alpha-1}{\alpha-2} k_\mathrm{min} 
	\propto k_{\mathrm{min}}.
\end{equation}
Therefore, the ratio of the number of links in the coarse-grained network to the original one is 
\begin{equation}
	\frac{L'}{L} = \frac{n'\langle k' \rangle}{n\langle k\rangle} =\frac{n'}{n} \frac{k'_{\mathrm{min}}}{k_{\mathrm{min}}} = 2^{\frac{2}{\alpha-1}-d_k-2},
\end{equation}
which leads to the link exponent 
\begin{equation}
	d_L =-\frac{\ln 	\frac{L'}{L}}{\ln 	\frac{n}{n'}} = 1-\frac{1}{\alpha-1}+\frac{d_k}{2}.
\end{equation}
To link $d_L$ back to the structural parameters of the original network (the finest-grained level), we utilize the following relation to replace the index $d_k$ with the box-covering dimension $d_B$:
\begin{equation}
	\alpha = 1 + \frac{d_B}{d_k}.
\end{equation}
$d_B$ can be numerically determined in the original network with a well-established box-covering algorithm~\cite{Song2007, Song2005,Kim2007}. The final expression of the link exponent is:
\begin{equation}
	d_L  = 1-\frac{1}{\alpha-1}+\frac{d_B}{2(\alpha-1)}=1+\frac{d_B-2}{2\qty(\alpha-1)},
\end{equation}
recovering Eq.~11 of the main text.

\section{Supplementary simulation results}
This section supplies additional numeric results that further support discussion in the main text and rationalize the approximations used in the analytical derivation. 

\subsection{The square lattice}
\subsubsection{Finite size and boundary effects}
\begin{figure}
	\centering
	\includegraphics[width=\linewidth]{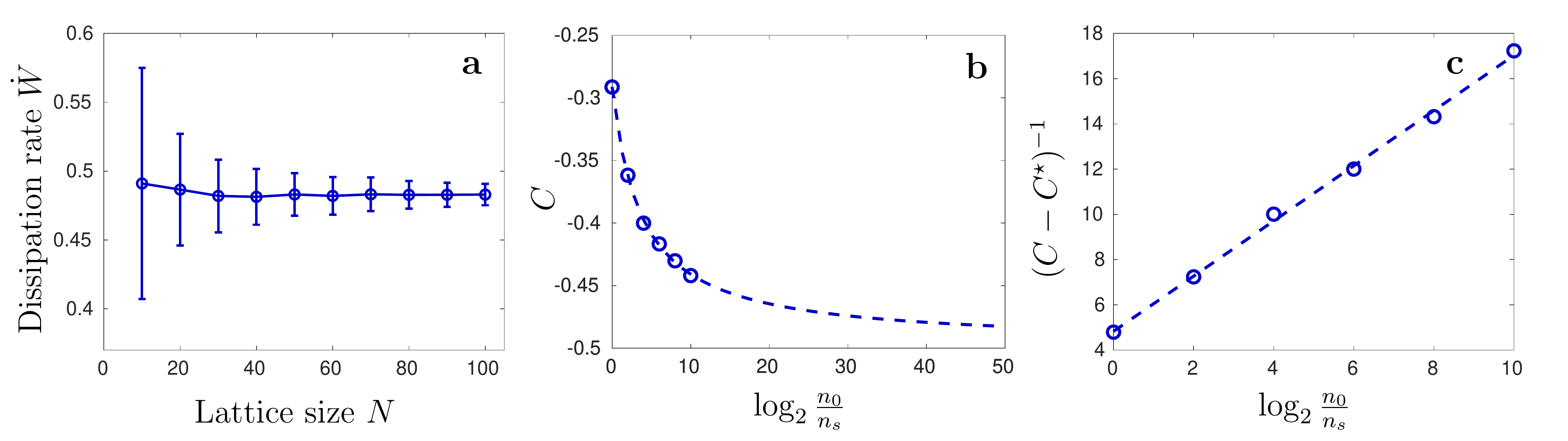}
	\caption{(a) The steady-state dissipation rate for square lattices with different side length $N$. Error bar: one standard deviation. The parameters used to generate the lognormal distribution of the transition rates $\ln k \sim \mathcal{N}(\mu,\sigma)$ are $\mu=0$, $\sigma=0.5$. (b) The convergence of correlation $C$ in the square lattice. Dashed line: numeric fit which extrapolates to the fixed point $C^\star=-0.5$. (c) To make the convergence in (b) more salient, we plot $\frac{1}{C-C^\star}$ against the block size. Its linear increase demonstrates a first-order convergence to $C^\star=-0.5$ (also see discussion in the SI text). }
	\label{Fig: finite size and correlation in 2d}
\end{figure}

The scaling behavior analyzed above assumes an infinitely large system which can approach a RG fixed point after infinite rounds of coarse-graining. In reality, nevertheless, we can only start with a finite system and do finite rounds of renormalization. Fortunately, the total dissipation rate does not depend on the lattice size $N$ as long as the total probability is normalized to $1$ and periodic boundary conditions are imposed on all sides (Fig.~\ref{Fig: finite size and correlation in 2d}a).

To evaluate the flux correlation $C^\star$ at the RG fixed point, we need to extrapolate the correlation from finite rounds of coarse-graining to the limit $s\to\infty$. As shown in Fig.~\ref{Fig: finite size and correlation in 2d}b, $C$ decreases with the block size following the relation:
\begin{equation}
	C\qty(\frac{n_0}{n_s}) =   \frac{x+p}{x+q} C^\star, \quad x = \log_2\frac{n_0}{n_s},
\end{equation}
with fitting parameters $C^\star =-0.50$, $p=2.19$, $q=3.74$, and $R^2=0.9992$. Apparently, $C$ converges to $C^\star$ at the RG fixed point. To make the convergence more salient, we demonstrate in Fig.~\ref{Fig: finite size and correlation in 2d}c that the inverse of the distance between $C$ and $C^\star$ increases linearly with $x=\log_2\frac{n_0}{n_s}$ and will eventually go to infinity. This correspond to first-order convergence with respect to $x$.

\subsubsection{Statistics of various quantities and justification for approximations}
It was demonstrated in the main text that the transition rates at all coarse-grained levels all follow lognormal distribution, which manifests the system's self-similarity. Here we present similar evidences concerning the local fluxes and dissipation rates of the system to further demonstrate the self-similar property. 

\begin{figure}
	\centering
		\includegraphics[width=\linewidth]{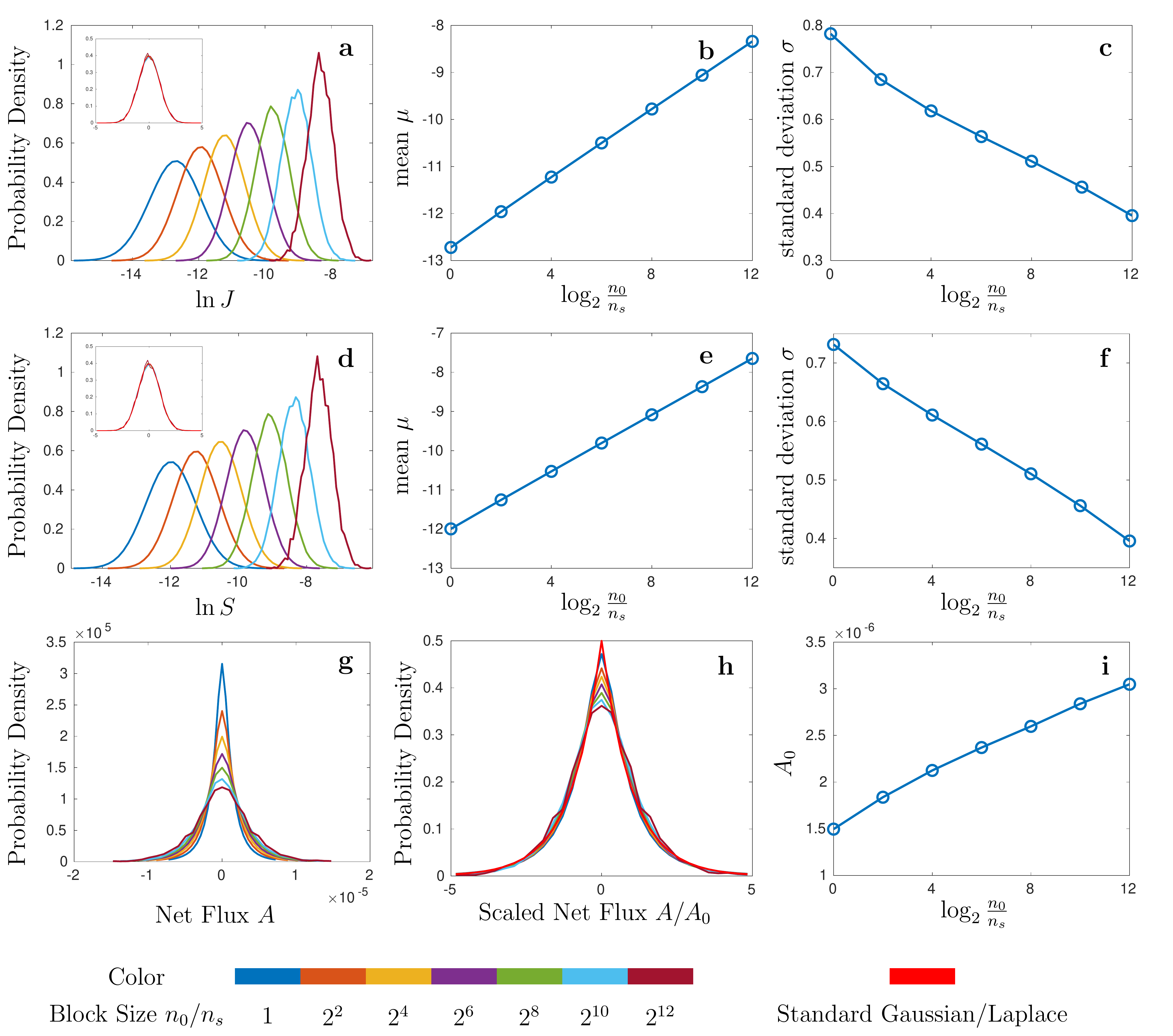}
	\caption{Statistics of various fluxes ($J$, $S$, $A$) at different coarse-grained levels. (a) The probability density function for $\ln J$. Inset: the distributions at all levels collapse to a standard Gaussian distribution $p(x) = \frac{1}{\sqrt{2\pi}}\exp\qty(-x^2/2)$ (red curve) after shifting the mean to $0$ and scaling the standard deviation to $1$. (b)--(c) The parameters $\mu$ and $\sigma$ of the lognormal distribution $\ln J \sim \mathcal{N}(\mu,\sigma)$ as a function of the block size $n_0/n_s$. (d)--(f) depicts the statistics of $\ln S$ in the exact same format as (a)--(c). (g) The probability density function for the net flux $\ln A$. (h) the distributions of the fluxes scaled by $A_0$. The red curve is Laplace distribution $p(x)=\frac{1}{2}\exp(-\qty|x|)$. (i) The distribution parameter $A_0$ as a function of the block size $n_0/n_s$.
	The corresponding relation between the colors in histograms (a), (d), (g), (h) and the block sizes of coarse-grained levels are given at the bottom. }
	\label{Fig: square lattice flux statistics J, S, A}
\end{figure}

Let us start by looking at the fluxes. Similar to the transition rates discussed in Fig.~2 of the main text, the transition fluxes $J$ also follow lognormal distributions at all coarse-grained levels with an increasing mean and decreasing standard deviation (Fig.~\ref{Fig: square lattice flux statistics J, S, A}a--c). Remarkably, the histograms of normalized $\ln J$ all collapse perfectly to a standard Gaussian distribution \ref{Fig: square lattice flux statistics J, S, A}a inset), namely
\begin{equation}
	\frac{\ln{J} - \mu}{\sigma} \sim \mathcal{N}\qty(0,1).
\end{equation}
The same behaviors are observed in the symmetric flux $S_{i,j} = J_{i,j} + J_{j,i}$ (Fig.~\ref{Fig: square lattice flux statistics J, S, A}d--f), including the collapsing of the normalized $\ln{S}$ to a standard Gaussian distribution (Fig.~\ref{Fig: square lattice flux statistics J, S, A}d inset). This is not completely surprising since the sum of two random variables independently drawn from identical lognormal distirbutions is approximately lognormal-distributed, per Fenton-Wilkinson approximation~\cite{Fenton1960,Mitchell1968}. In contrast, the antisymmetric flux $A$ no longer follows lognormal distribution. As demonstrated in Fig.~\ref{Fig: square lattice flux statistics J, S, A}g, it is symmetrically distributed with respect to $A=0$ and approximately decays exponentially on both sides. The probability density function is the Laplace distribution:
\begin{equation}
	p(A)=\frac{1}{2A_0}\exp(-\qty|A|/A_0).
\end{equation}
The parameter $A_0$ can be determined by max-likelihood estimation $\hat{A_0} = \langle \qty|A|\rangle$. It increases nearly linearly with the logarithm of the block size $\log_2\qty(n_0/n_s)$ (Fig.~\ref{Fig: square lattice flux statistics J, S, A}i). The distribution of the scaled net fluxes $A/A_0$ at all coarse-grained levels collapse to a (unit) Laplace distribution $p(x)=\frac{1}{2}\exp(-\qty|x|)$, except for deviations near $x=0$ (Fig.~\ref{Fig: square lattice flux statistics J, S, A}h).

Notably, the growth of $\mu$ and $A_0$ for the fluxes in Fig.~\ref{Fig: square lattice flux statistics J, S, A} entails that $S$ grows as fast as the block size $n_0/n_s$ while $\qty|A|$ merely grows as fast as the logarithm of the block size, namely $\qty|A|\sim \log_2\qty(n_0/n_s)$. Therefore, $\qty|A/S|$ must decay to zero as the block size goes to infinity, further justifying the approximation  $\qty|A/S|\ll 1$ employed in Eq.~\ref{Equ: total dissipation in A and S}. 

\begin{figure}
	\centering
	\includegraphics[width=.4\linewidth]{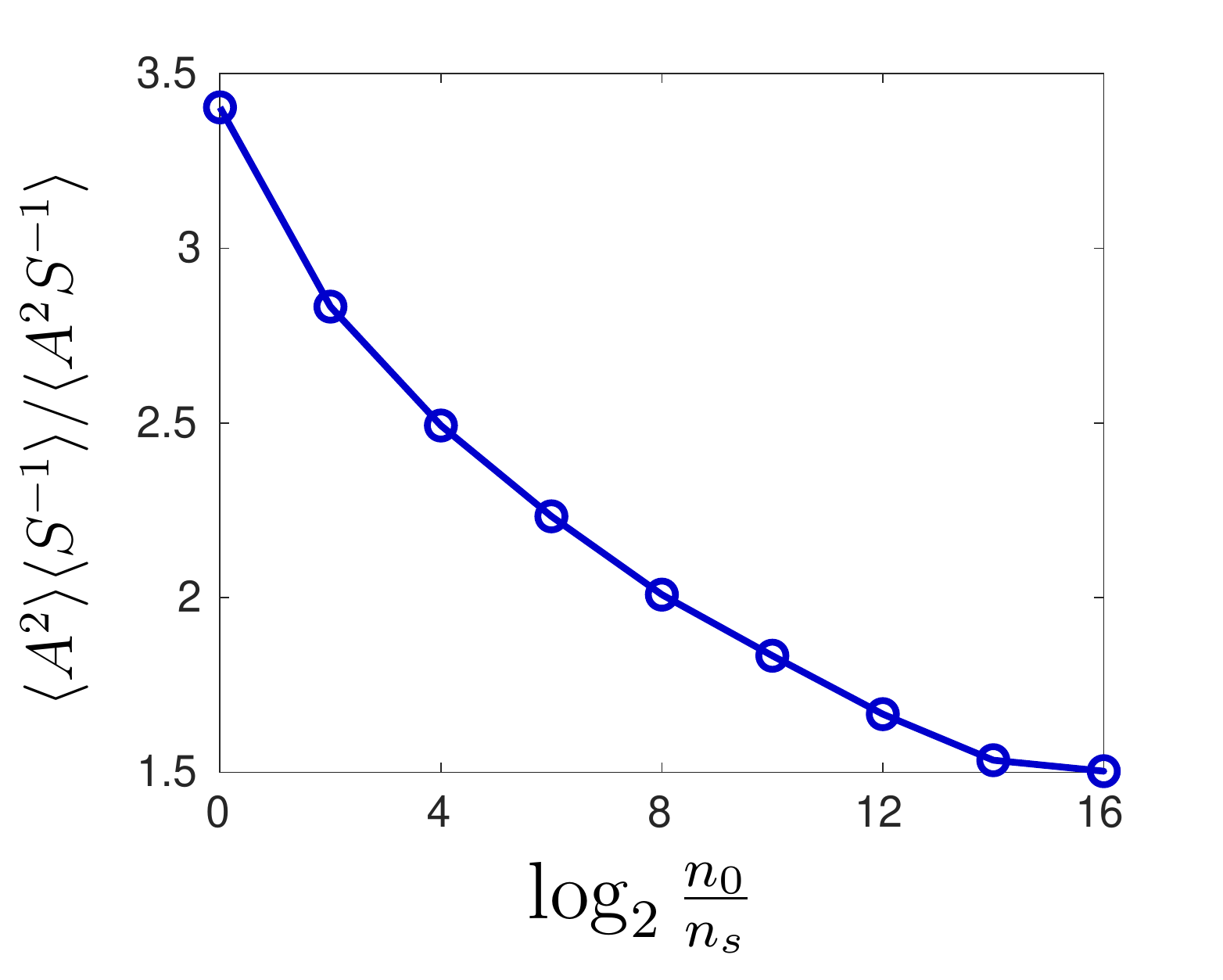}
	\caption{The ratio of $\langle A^2 \rangle\langle S^{-1} \rangle$ to $\langle A^2 S^{-1} \rangle$ as a function of the block size $n_0/n_s$. }
	\label{Fig: square lattice flux correlation }
\end{figure}

We also numerically evaluate the validity of approximations taken in Eq.~\ref{Equ: A2 S approximation in the main derivation}. Concretely, it includes two layers of approximations: 
\begin{equation}
\frac{\left\langle  \frac{A_{i,j}^2}{S_{i,j}} \right\rangle}{\left\langle  \frac{A_{i_\alpha,j_\beta}^2}{S_{i_\alpha,j_\beta}} \right\rangle} 
\approx \frac{\left\langle {A_{i,j}^2} \right\rangle\left\langle S_{i,j}^{-1} \right\rangle}{\left\langle  {A_{i_\alpha,j_\beta}^2} \right\rangle \left\langle {S_{i_\alpha,j_\beta}^{-1}} \right\rangle}
\approx \frac{\langle {A_{i,j}^2} \rangle}{\langle  {A_{i_\alpha,j_\beta}^2} \rangle} \frac{\langle S_{i_\alpha,j_\beta}\rangle}{\langle S_{i,j}\rangle} = 1+C
\end{equation}
The first approximation concerns the correlation between $A^2$ and $S^{-1}$. The second approximation concerns the difference between $1/\langle S \rangle $ and $\langle S^{-1} \rangle$. For the first one, we directly calculate the ratio of $\langle A^2 \rangle\langle S^{-1} \rangle$ to $\langle A^2 S^{-1} \rangle$ and plot it in Fig.~\ref{Fig: square lattice flux correlation }. The deviation from $1$ indicates negative correlation between $A^2$ and $S^{-1}$. The ratio effectively adds another term to the dissipation scaling exponent. The magnitude of change to the scaling exponent is 
\begin{equation}
\delta \lambda  = \frac{ \Delta\log_2\qty( \langle A^2 \rangle\langle S^{-1} \rangle/\langle A^2 S^{-1} \rangle)}{\Delta \log_2\qty(n_0/n_{s})} = \frac{\ln\qty(3.40/1.54)}{16\ln 2} = 0.07,
\end{equation}
which is small compared to the $\log_4(1+C)$ term due to correlation. 

Now we turn to the second approximation. Since we know that $S$ follows a lognormal distribution $\mathcal{N}\qty(\mu,\sigma)$, the difference between $1/\langle S\rangle$ and $\langle S^{-1} \rangle$ can be computed analytically:
\begin{equation}
	\begin{aligned}
		\frac{1}{\langle S \rangle} &= \qty[\int_{0}^{+\infty}\frac{1}{\sqrt{2\pi\sigma^2}} \exp\qty(-\frac{\qty(\ln x-\mu)}{2\sigma^2}) \dd{x}]^{-1} =  e^{-\mu -\frac{\sigma ^2}{2}} , \\
		\langle \frac{1}{S} \rangle &= \int_{0}^{+\infty}\frac{1}{\sqrt{2\pi\sigma^2}}\exp\qty(-\frac{\qty(\ln x-\mu)}{2\sigma^2}) \frac{\dd{x}}{x^2} = e^{\frac{\sigma ^2}{2}-\mu }.
	\end{aligned}
\end{equation}
The ratio is 
\begin{equation}
	\langle \frac{1}{S} \rangle/\frac{1}{\langle S \rangle} = e^{\sigma^2}
\end{equation}
Therefore, the second approximation 
\begin{equation}
	\frac{\frac{\left\langle {A_{i,j}^2} \right\rangle\left\langle S_{i,j}^{-1} \right\rangle}{\left\langle  {A_{i_\alpha,j_\beta}^2} \right\rangle \left\langle {S_{i_\alpha,j_\beta}^{-1}} \right\rangle}
	 }{\frac{\langle {A_{i,j}^2} \rangle}{\langle  {A_{i_\alpha,j_\beta}^2} \rangle} \frac{\langle S_{i_\alpha,j_\beta}\rangle}{\langle S_{i,j}\rangle}} = \exp\qty(\sigma_s^2-\sigma_{s-1}^2).
\end{equation}
The change of $\sigma^2$ can be estimated from Fig.~\ref{Fig: square lattice flux statistics J, S, A}f. $\sigma$ decreases from $0.73$ to $0.40$ as the block size increases from $1$ to $2^{12}$. Therefore, the influence of this approximation to the scaling exponent is 
\begin{equation}
	\delta \lambda  = \frac{\log_2e^{\sigma_0^2-\sigma_{12}^2}}{\log_2\qty(n_0/n_{12})} = \frac{0.73^2-0.40^2}{12\ln 2} = 0.04,
\end{equation}
which is negligible. 

\begin{figure}
	\centering
	\includegraphics[width=.8\linewidth]{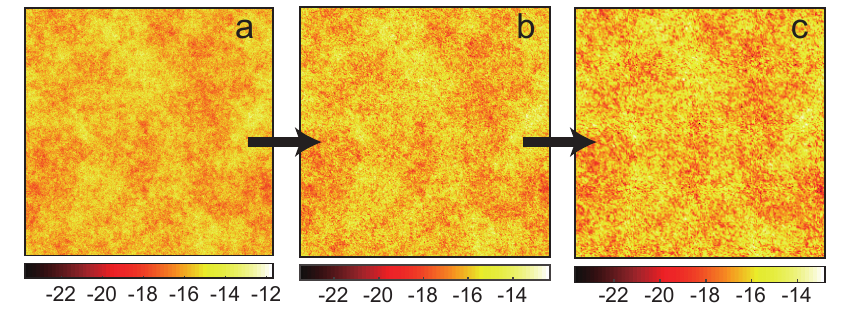}
	\caption{Local dissipation rate profile in the square lattice (in log scale, plotting $\ln w_{i,j}$). (a) The original network of size $N=1024$. The transition rates follow a lognormal distribution with mean $\mu=0$ and standard deviation $\sigma=0.5$. (b) The lattice resulting from the coarse-graining of (a), of size $N=512$. (c) The lattice resulting from the coarse-graining of (b), of size $N=256$. }
	\label{Fig: square lattice dissipation profile}
\end{figure}

\subsection{The spatial profile of dissipation rate}
Next, we investigate the spatial profile of dissipation. The local dissipation rate at site $(i,j)$ is defined by the average of the dissipation in all four links associated with it, namely
\begin{equation}
	w_{i,j} = \frac{1}{4} \sum_{\alpha=1}^{4} \qty(J_{(i,j)\to (i_\alpha,j_\alpha)}-J_{(i_\alpha,j_\alpha)\to (i,j)}) \ln\frac{J_{(i,j)\to (i_\alpha,j_\alpha)}}{J_{(i_\alpha,j_\alpha)\to (i,j)}}.
\end{equation}
$(i_\alpha,j_\alpha)$ ($\alpha=1,2,3,4$) are the four neighbors adjacent to $(i,j)$, namely $(i,j+1)$, $(i,j-1)$, $(i+1,j)$, and $(i-1,j)$.
The local dissipation rate is self-similar in the sense of both its spatial profile and overall distribution.
Fig.~\ref{Fig: square lattice dissipation profile} demonstrates the spatial profile of $\ln w_{i,j}$ at the first three levels. The arrows indicate the direction of coarse-graining.  The overall profiles are similar. 

\begin{figure}
	\centering
	\includegraphics[width=.8\linewidth]{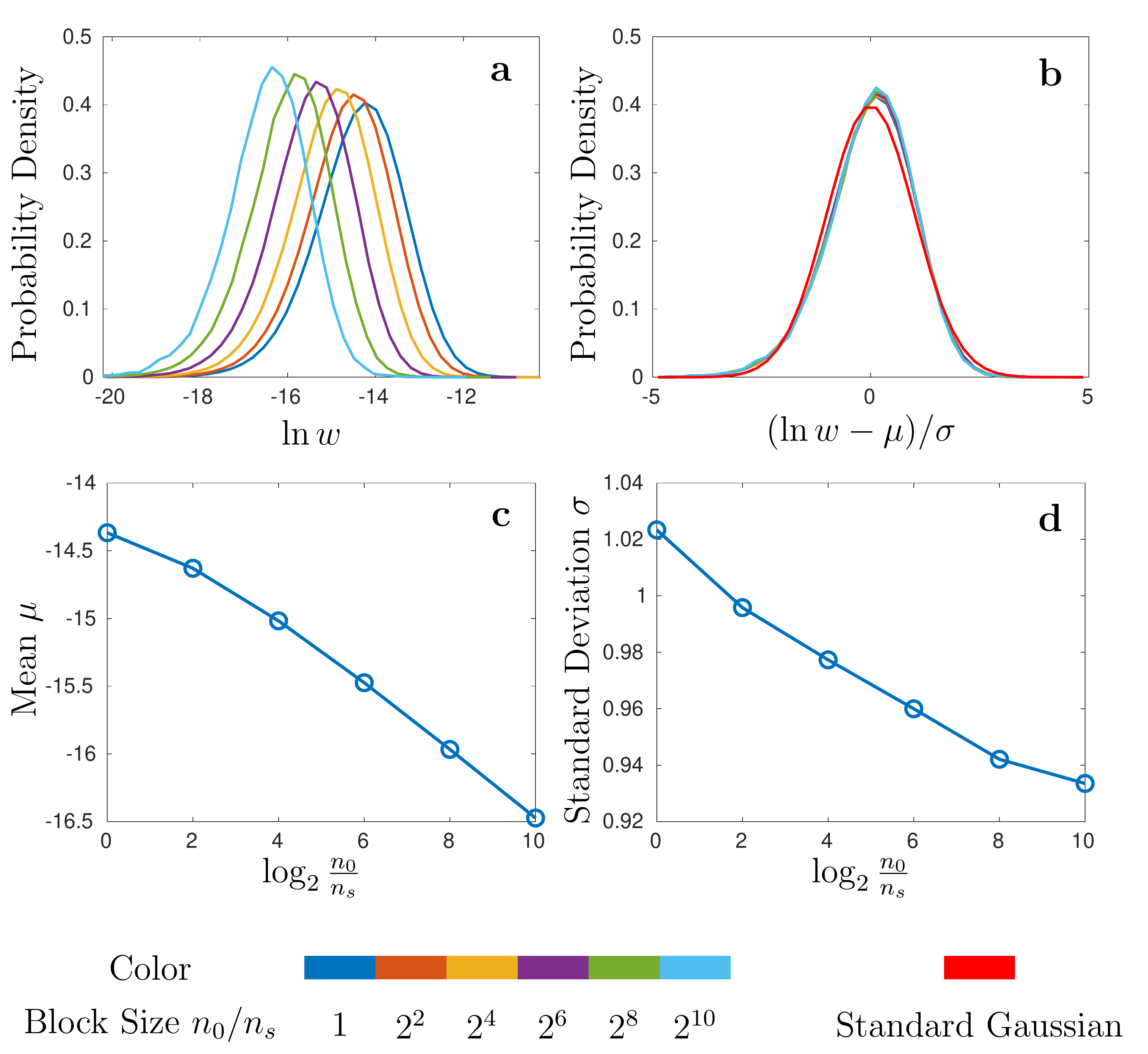}
	\caption{Statistics of the local dissipation rate $w_{i,j}$ (in log scale). (a) The probability density function of $\ln w$ at different coarse-grained levels of the network (see the bottom of the figure for color denotation). (b) After shifted to zero mean and scaled to unit standard deviation, distributions of $\ln{w}$ at all levels collapse to a single distribution. However, this distribution deviates from the standard Gaussian distribution (red curve). (c)--(d) mean $\mu$ and $\sigma$ of $\ln w$ at different coarse-grained levels, as functions of the block size.  }
	\label{Fig: square lattice dissipation statistics }
\end{figure}

Fig.~\ref{Fig: square lattice dissipation statistics } presents the statistical distribution of $\ln w_{i,j}$. The distribution at all scales collapse to a single distribution function after shifting the mean to zero and scaling the standard deviation to one (Fig.~\ref{Fig: square lattice dissipation statistics }b), which stands as another example of self-similarity. Unlike the common distribution function for $\ln J$, which is Gaussian, the one found here undoubtedly deviates from the standard Gaussian distribution. The mean and standard deviation of $\ln w$ are presented in Fig.~\ref{Fig: square lattice dissipation statistics }c--d.

\subsection{Dependence on the initial rate distribution}

\begin{figure}
	\centering
	\includegraphics[width=\linewidth]{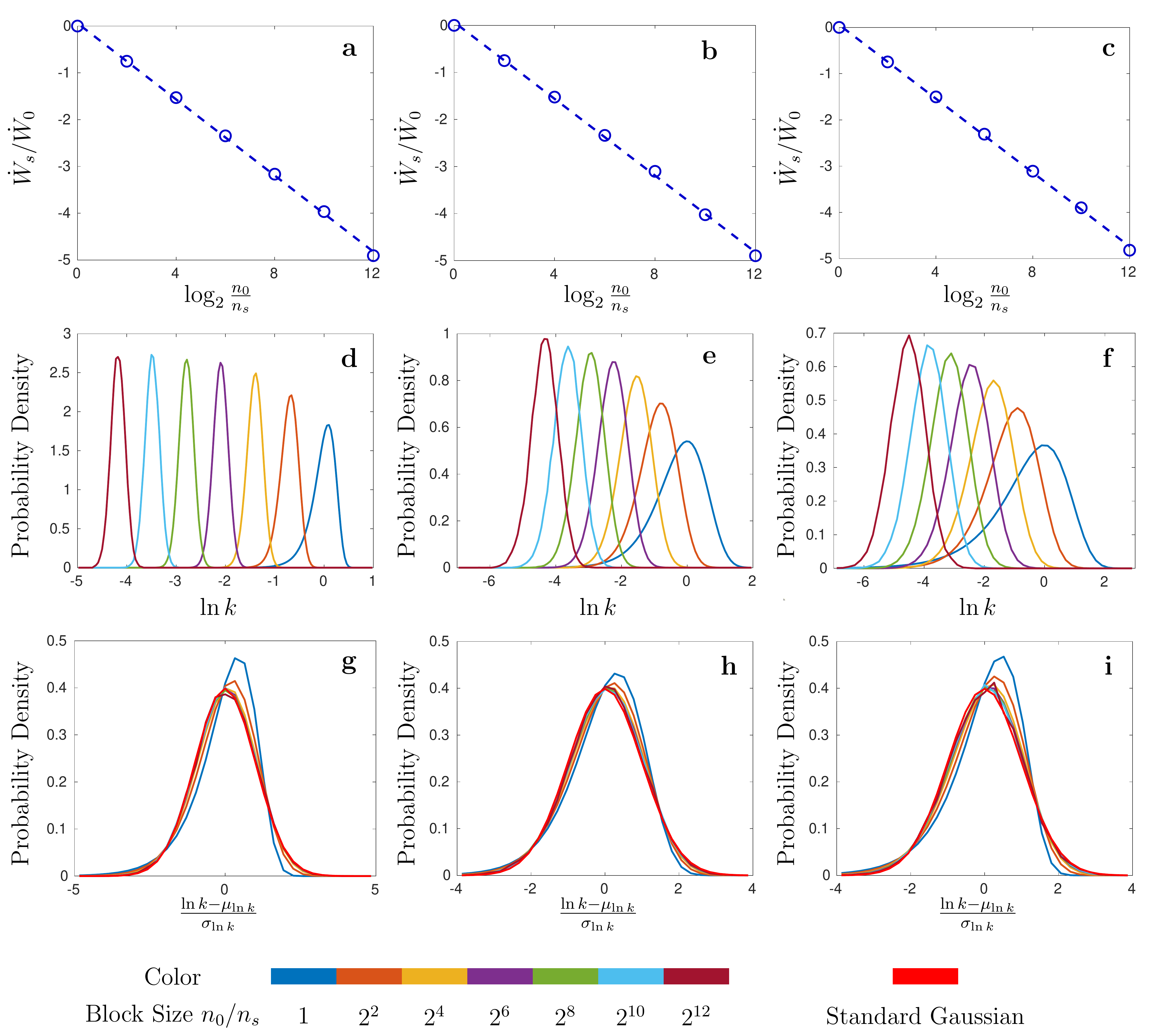}
	\caption{Results  in the square lattice where transitions follow Weibull distribution (left panel), gamma distribution (middle panel), or exponential distribution (right panel). (a)--(c): The dissipation rate decreases in a power-law relation with the block size. The scaling exponents from the least-square fit are $1.35$, $1.35$, and $1.33$. (d)--(f): The distribution of $\ln{k}$ ($k$ is the transition rate) at different coarse-grained levels. (g)--(f) The distribution of $\frac{\ln k -\mu_{\ln k} }{\sigma_{\ln k}}$ at different coarse-grained levels, where $\mu_{\ln{k}}$ and $\sigma_{\ln k}$ are the mean and the standard deviation of $\ln{k}$, respectively. The transformation shifts the mean to zero and scales the standard deviation to $1$ to allow the comparison between the shape of the probability density functions. The red curve is the standard Guassian distribution $p(x) = \frac{1}{\sqrt{2\pi}}\exp\qty(-x^2/2)$.  The corresponding relation between the colors in histograms (d)--(i) and the block sizes of coarse-grained levels are given at the bottom. }
	\label{Fig: square lattice other distributions}
\end{figure}

It is apparent from the derivation in section IA of the SI that the dissipation scaling relation does not depend on the exact form of initial transition rate distribution. To demonstrate this point, we performed all the studies in the square lattice with transition rates sampled from three other distributions: Weibull distribution, gamma distribution, and exponential distribution. The probability density function for these distributions are: 
\begin{equation}
\begin{aligned}
	\text{Weibull: } &  f(x;\lambda,k) = \frac{k}{\lambda} \qty(\frac{x}{\lambda})^{k-1} \exp\qty(-\qty(\frac{x}{\lambda})^k) & \quad \qty(x > 0)\\
	\text{gamma: } & f(x;\alpha,\beta) = \frac{\beta^\alpha}{\Gamma(\alpha)} x^{\alpha-1} e^{-\beta x} & \quad \qty(x > 0)\\
	\text{exponential: } & f(x;\mu) = \frac{1}{\mu} \exp\qty(-\frac{x}{\mu}) & \quad \qty(x > 0)
\end{aligned}
\end{equation}
The last one is a special case of the gamma distribution (with $\alpha=1$).

Fig.~\ref{Fig: square lattice other distributions} presents the results with these three distributions. The parameters are chosen as
\begin{equation}
	k = 5, \lambda = \frac{1}{\Gamma\qty(1+1/k)}; \enspace \alpha = \beta = 2; \enspace \mu=1,
\end{equation}
such that the mean of the transition rates is $1$ in all three cases. In all three systems, the dissipation scaling behaviors are identical to that displayed in Fig.~2c of the main text, with scaling exponents close to 1.35 (Fig.~\ref{Fig: square lattice other distributions}a--c). However, the distribution of transition rates evolves differently (Fig.~\ref{Fig: square lattice other distributions}d--f). If the original transition rate distribution is lognormal, it stays lognormal at all coarse-grained levels (Fig.~2a, main text). On the contrary, the distributions here start with a considerable deviation from lognormal at the finest level but gradually converge to lognormal as the system is coarse grained (Fig.~\ref{Fig: square lattice other distributions}g--i). In other words, the lognormal distribution is an stable attractor for the transition rates distribution, i.e. the rates distribution at the RG fixed point. Since we are mainly concerned with the scaling properties at the RG fixed point of the reaction network, it is reasonable to start with a lognormal distribution and study the convergence or scaling of other properties as the system is coarse-grained. Therefore, it suffices to consider only lognormal-distributed rates in all networks studied here.

\subsection{The random-hierarchical network}

Table \ref{Tab: random hierarchical network results} presents the scaling exponents for different random hierarchical networks constructed with parameters $d$ (the average degree) and $m$ (the average cluster size). The last column shows that the dissipation scaling exponent $\lambda$ is in close proximity to $1$ regardless of $d$ and $m$, corroborating the conclusion $\lambda_\mathrm{RHN}=1$ in the main text. 

\begin{table}
	\centering
	\begin{ruledtabular}
		\begin{tabular}{ c c c }
			$d$ & $m$  & $\lambda$ \\ 
			\hline  
			4 & 4 & $1.057 \pm 0.027$ \\
			6 & 4 &  $1.068 \pm 0.021$  \\ 
			6 & 6 & $1.070 \pm 0.022$  \\ 
			8 & 6 &  $1.075 \pm 0.020$  \\ 
			10 & 6  & $1.077 \pm 0.021$ \\ 
			8 & 8 & $1.055 \pm 0.015$ \\
		\end{tabular}
	\end{ruledtabular}
	\caption{The dissipation scaling exponents of the random hierarchical network (mean $\pm$ one standard deviation, over 100 replica). The scaling exponent $\lambda$ is close to $1$, independent of the average degree $d$ and the number of states with each macroscopic state $m$.}
	\label{Tab: random hierarchical network results}
\end{table}

\subsection{The scale-free network}

\begin{figure}
	\centering
	\includegraphics[width=.8\linewidth]{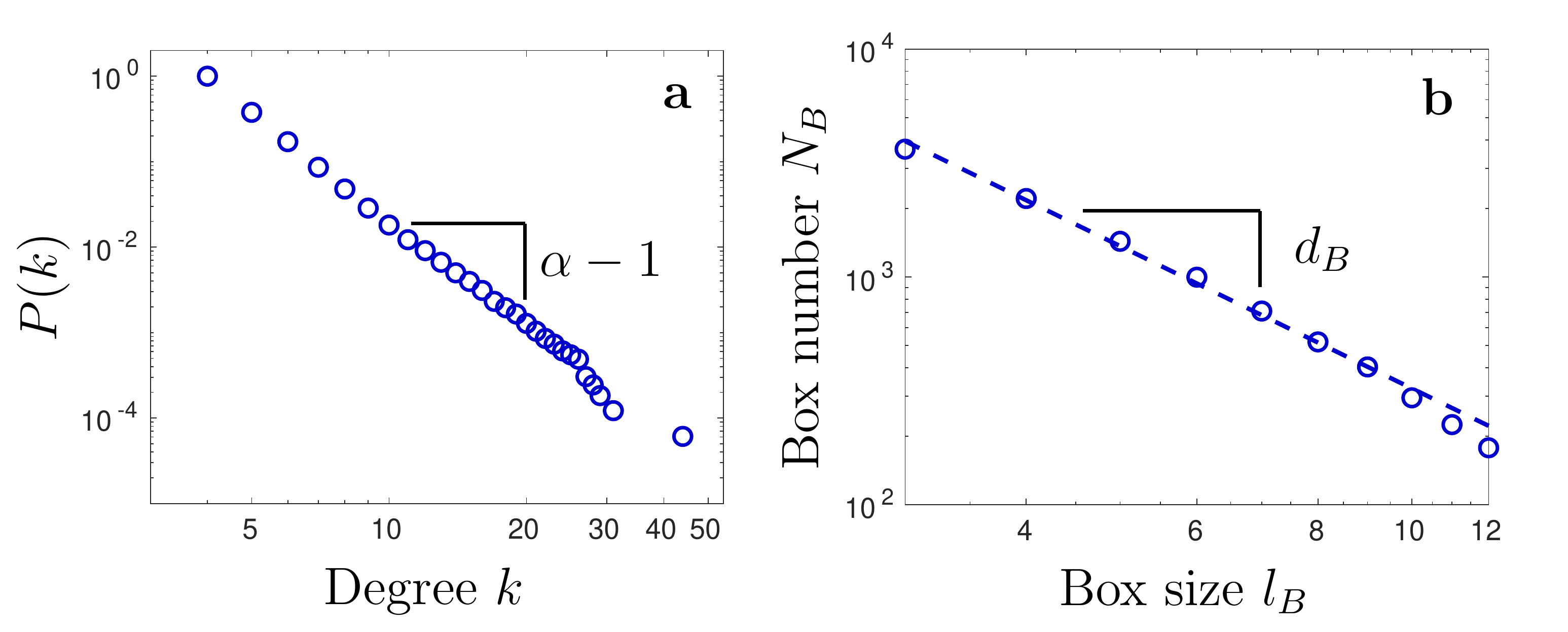}
	\caption{(a) The cumulative degree distribution is power-law ($P(k)\propto k^{1-\alpha}$). The slope is $\alpha-1=4.00$ in this particular example. (b) The power-law relation between $N_B$ and $l_B$ reveals a fractal dimension $d_B$. The slope is $d_B=1.98$ in this particular example. }
	\label{Fig: scale-free network gamma and dB fitting}
\end{figure}

We first demonstrate the determination of $\alpha$ and $d_B$ with the example in Fig.~\ref{Fig: scale-free network gamma and dB fitting}. After creating the scale-free network with the method described in the main text and refs.~\cite{Rozenfeld2002,Kim2004}, we plot the cumulative degree distribution $P(k)\sim \qty(\frac{k_\mathrm{min}}{k})^{\alpha-1}$ (see Eq.~\ref{Equ: scale-free CDF}) against $k$ on a log-log scale and determine the slope $\alpha-1$ by linear fitting and therefore determine $\alpha$. The index $d_B$ is obtained similarly by fitting the power-law relation between the box size $l_B$ and the minimum number of boxes containing states whose pairwise distance does not exceed $l_B$ that are needed to cover the entire graph, $N_B$. Since finding the exact minimum $N_B(l_B)$ is NP-hard, we instead use a greedy coloring algorithm that is known to provide good approximations to the value of $N_B(l_B)$~\cite{Song2007}.

\subsection{Networks without the dissipation scaling relation}
Finally, we introduce two networks in which the dissipation does not obey a power-law relation with respect to the block size after coarse-graining and discuss why this scaling relation breaks down. 

\subsubsection{The Erd{\H{o}}s-R{\'{e}}nyi random network}

\begin{figure}
	\centering
		\includegraphics[width=.8\linewidth]{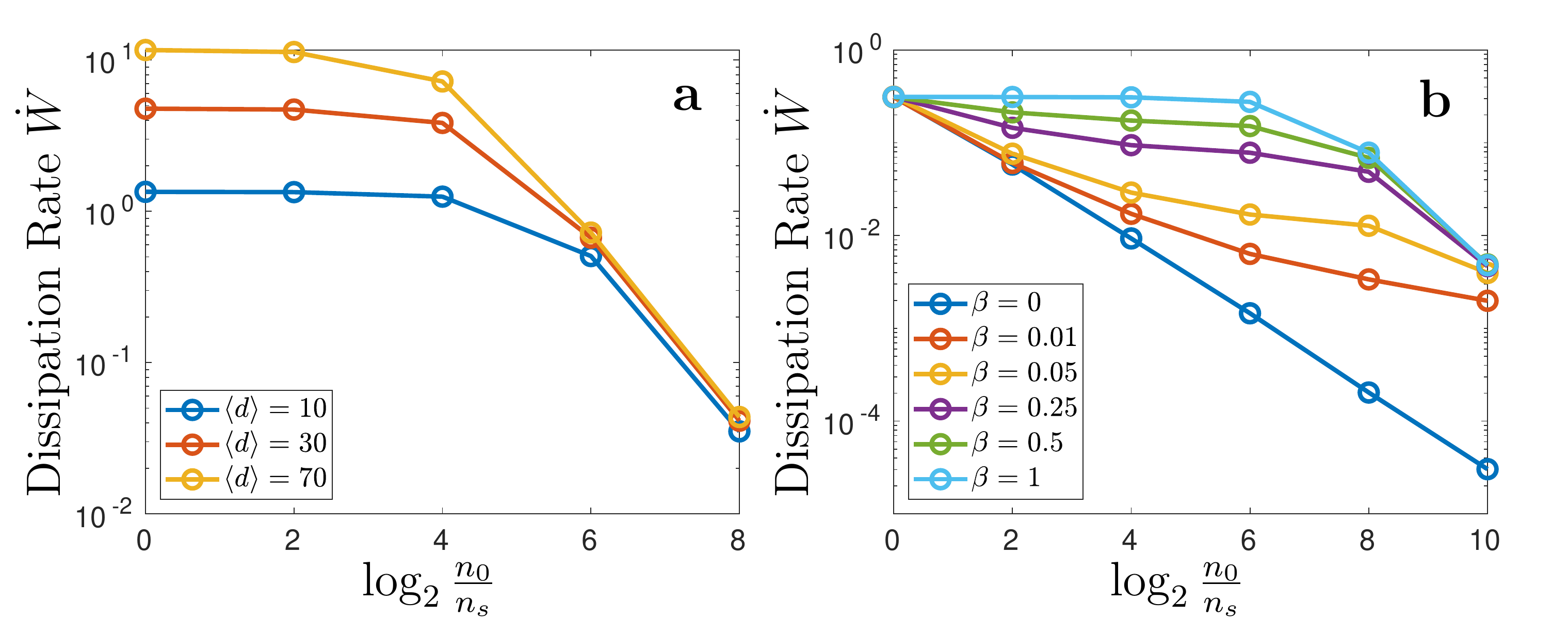}
	\caption{The dissipation rate in the Erd{\H{o}}s-R{\'{e}}nyi random network and Watts-Strogatz small-world network. The initial transition rates are drawn from lognormal distribution $\ln k \sim \mathcal{N} (\mu,\sigma)$ with $\mu=0$ and $\sigma=0.4$. (a) Results in Erd{\H{o}}s-R{\'{e}}nyi random network with different average degrees $\langle d\rangle$ for $N_0\times N_0$ nodes ($N_0=128$). (b) Results in the Watts-Strogatz small-world network with different rewiring probabilities $\beta$ (square lattice with $256\times 256$ nodes).}
	\label{Fig: ER network}
\end{figure}

We create an Erd{\H{o}}s-R{\'{e}}nyi random network with a binomial model that is equivalent to the method used in the classic paper by Erd{\H{o}}s and R{\'{e}}nyi~\cite{Albert2002}. To generate a network with average degree $d$, we connect each pair of states on a $N_0$ by $N_0$ square lattice with a constant probability $p = \frac{d}{N_0^2(N_0^2-1)}$. The coarse-graining is conducted in the exact same way as that used in the 2d-embedded scale-free network.  We find that the dissipation decreases with the block size but does not follow a power-law relation (Fig.~\ref{Fig: ER network}a). This is not unexpected since the coarse-graining completely disregards the connectivity of the network. In the scale-free network, states are preferentially linked to neighbors by construction, which relates the network topology with the geography of the square lattice. Hence, grouping together geographically adjacent states during coarse-graining also means that states that are more interconnected (i.e., neighbors) are more likely to belong to the same macrostate. In stark contrast, a state in an Erd{\H{o}}s-R{\'{e}}nyi random network has equal probabilities of being connected to a state that belong to the same macrostate or a state that does not. Therefore, the first few rounds of coarse-graining of the sparse network almost does not involve any link merger, while the system approaches a complete graph at later rounds of coarse-graining with the number of links rapidly decreasing. The discrepancy between early and later stages of coarse-graining reveals the lack of self-similarity in the Erd{\H{o}}s-R{\'{e}}nyi random network, which inevitably leads to the absence of dissipation scaling. 

\subsubsection{The Watts-Strogatz small-world network}

The Watts-Strogatz network can be created by randomly rewiring the links in a regular square lattice with probability $\beta\in(0,1)$~\cite{Watts1998}. Coarse-graining can be carried out in the same fashion as scale-free network and Erd{\H{o}}s-R{\'{e}}nyi random network.  The network approaches a square lattice as $\beta\to0$ and approaches the Erd{\H{o}}s-R{\'{e}}nyi random network as $\beta\to 1$. Therefore, we expect the dissipation to scale with the block size as $\beta$ tends to 0 and to breaks down as $\beta$ is increased.  Indeed, this is what we find numerically (Fig.~\ref{Fig: ER network}b). The same arguments used in the random network can be employed to account for the breakdown of the scaling property here. 

\section{Hidden free energy costs in the coarse-grained network}
In this section, we elucidate the physical origin of the difference between dissipation at fine-grained and coarse-grained levels of the network. For the sake of illustration, let's first consider the regular lattice in Fig.~1a of the main text. Let $\dot{W}_0$ be the dissipation rate at the finest-grained level. This dissipation rate can be decomposed into three components 
\begin{equation}
	\dot{W}_{0}=\dot{W}_{0}^{(1)}+\dot{W}_{0}^{(2)}+\dot{W}_{0}^{(3)},
	\label{Eq: W0 decompositions}
\end{equation}
all of which are nonnegative~\cite{Esposito2012}. The first term $\dot{W}_{0}^{(1)}$ represents the dissipation associated with the coarse-grained transitions between CG states, which is exactly the total dissipation rate of the first CG level $\dot{W}_1$:
\begin{equation}
	\dot{W}_{0}^{(1)} = \dot{W}_1  = 4^{-\lambda} \dot{W}_0 = \frac{1+C^*}{4} \dot{W}_0.
	\label{Eq: W0 first component}
\end{equation} 
The second term $\dot{W}_{0}^{(2)}$ is the dissipation caused by transitions within each CG state, which necessary for sustaining the internal nonequilibrium steady state. In the homogeneous lattice we study, this is proportional to the number of internal links:
\begin{equation}
	\dot{W}_{0}^{(2)} = \frac{L_\mathrm{internal}}{L_\mathrm{total}}\dot{W}_0 = \frac{1}{2}\dot{W}_0,
	\label{Eq: W0 second component}
\end{equation} 
where $L_\mathrm{internal}=N^2$ is the number of internal links and and $L_\mathrm{total}=2N^2$ is the total number of links in a regular lattice of size $N$. 

The third term $\dot{W}_{0}^{(3)}$ is associated with merging multiple (in this case, two) reaction pathways into one during coarse-graining. As shown in the example Fig.~1a in the main text, the two pathways $i_2 \leftrightarrow j_1$ and $i_4 \leftrightarrow j_3$ are merged into a single reaction $i\leftrightarrow j$. The randomness in which of the two microscopic transitions actually happened when a (macroscopic) transition between CG states $i$ and $j$ is observed contributes to this extra term of dissipation. Concretely, the total dissipation rate in the two microscopic pathways is
\begin{equation}
	\sigma_0(i,j)  = \qty(J_{i_2,j_1}-J_{j_1,i_2}) \ln \frac{J_{i_2,j_1}}{J_{j_1,i_2}} + \qty(J_{i_4,j_3}-J_{j_3,i_4}) \ln \frac{J_{i_4,j_3}}{J_{j_3,i_4}}.
\end{equation}
The total dissipation for the coarse-grained pathway is
\begin{equation}
	\sigma_1(i,j) =  \qty(J_{i_2,j_1}+J_{i_4,j_3}-J_{j_1,i_2}-J_{j_3,i_4}) \ln \frac{J_{i_2,j_1}+J_{i_4,j_3}}{J_{j_1,i_2}+J_{j_3,i_4}}.
\end{equation}
Their difference leads to $\dot{W}_{0}^{(3)}$. 
Following the same method in section~I.A.1 of the SI, we decompose the fluxes into symmetric and antisymmetric components to determine the average ratio of $\sigma_1$ to $\sigma_0$, which reads
\begin{equation}
	\left\langle \frac{\sigma_1(i,j)}{\sigma_0(i,j)} \right\rangle  = \frac{1}{2}
	\left\langle \frac{\qty(A_{i_2,j_1} + A_{i_4,j_3})^2}{A_{i_2,j_1}^2 + A_{i_4,j_3}^2} \right\rangle = \frac{1 + C^\star}{2}
\end{equation}
Since $C^\star<0$, the dissipation in the coarse-grained pathway is always smaller than the sum of dissipation in the two microscopic pathways. Their difference $\qty(\sigma_0(i,j)-\sigma_1(i,j))$ makes up the third term $\dot{W}_{0}^{(3)}$. We sum up such terms over the entire network:
\begin{equation}
   \dot{W}_{0}^{(3)} = \sum_{i<j}\qty(\sigma_0(i,j)-\sigma_1(i,j)).
\end{equation}
On the other hand, the total dissipation rate of microscopic transitions between different CG states is 
\begin{equation}
	\dot{W}_0 - \dot{W}_{0}^{(2)}= \frac{1}{2}\dot{W}_0 = \sum_{i<j}\sigma_0(i,j).
\end{equation}
Therefore $\dot{W}_{0}^{(3)}$ can be simplified to
\begin{equation}
	\dot{W}_{0}^{(3)} = \frac{1}{2}\dot{W}_0 \qty(1-\left\langle \frac{\sigma_1(i,j)}{\sigma_0(i,j)} \right\rangle ) = \frac{1-C^\star}{4}\dot{W}_0.
	\label{Eq: W0 third component}
\end{equation}
It can be verified that the sum of the three components in Eq.~\ref{Eq: W0 first component}, \ref{Eq: W0 second component} and \ref{Eq: W0 third component} exactly makes up the total dissipation $\dot{W}_0$ (Eq.~\ref{Eq: W0 decompositions}). 

The decomposition suggests that the dissipation that is missing at the coarse-grained levels consists of two components: $\dot{W}^{(2)}$, which arises from ignoring transitions within coarse-grained states, and $\dot{W}^{(3)}$, which stems from merging the transition pathways between coarse-grained states. The former maintains the internal nonequilibrium state of CG states, and the latter is associated with the multiplicity of ways in which a macroscopic transition can occur. Each of the microscopic pathways has a different nonequilibrium driving force $\gamma$. 

\begin{figure}
	\centering
	\includegraphics[width=.5\linewidth]{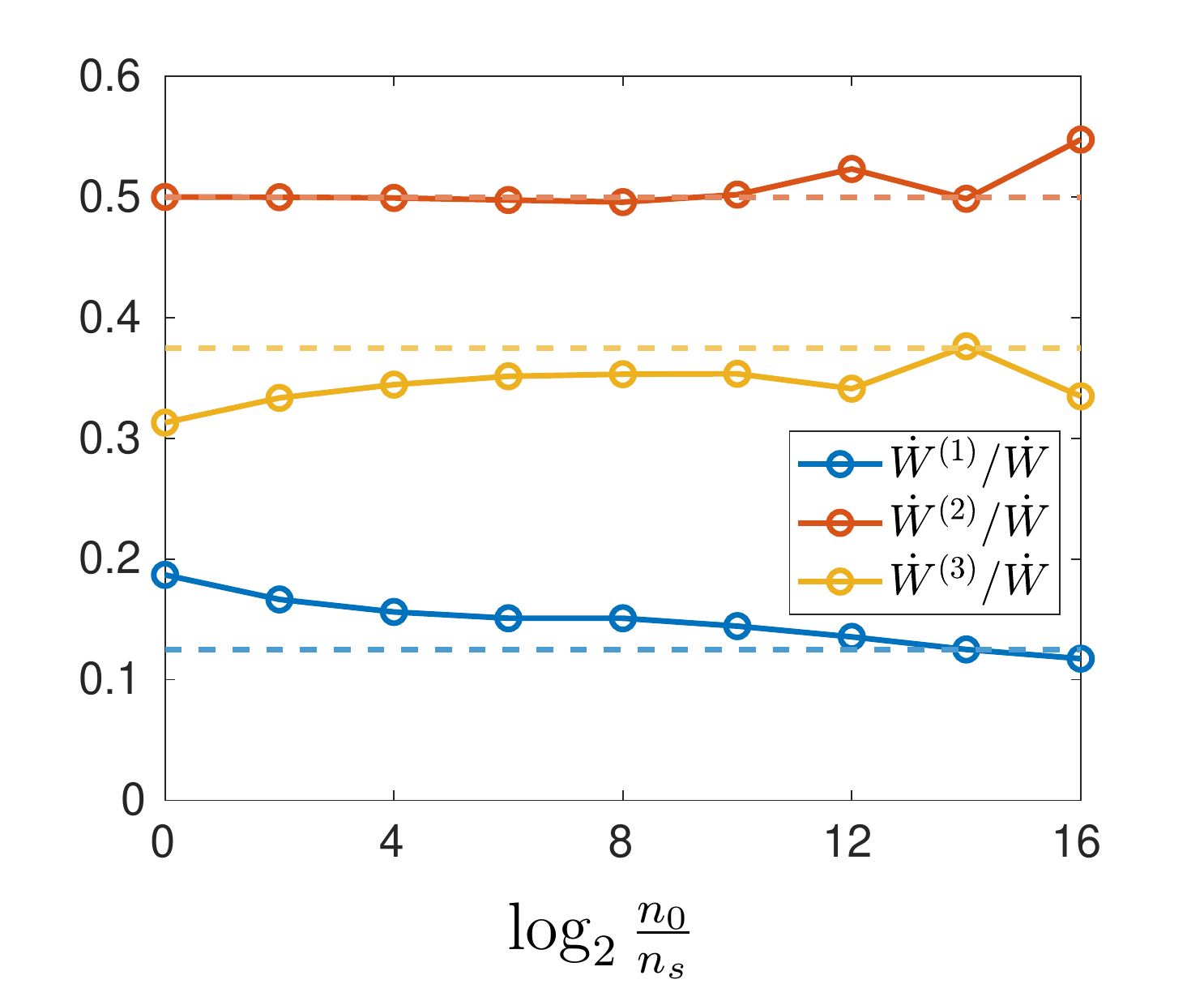}
	\caption{The partitioning ratios $\dot{W}^{(i)}/\dot{W}$ ($i=1,2,3$) in the square lattice. The circles are data points from simulation and the horizontal dashed lines are their expected values at the RG fixed point: blue, $\dot{W}^{(1)}/\dot{W}\to\frac{1+C^\star}{4}=0.125$ (Eq.~\ref{Eq: W0 first component}); orange, $\dot{W}^{(2)}/\dot{W}\to \frac{1}{2}$ (Eq.~\ref{Eq: W0 second component}); yellow, $\dot{W}^{(3)}/\dot{W}\to\frac{1-C^\star}{4}=0.375$ (Eq.~\ref{Eq: W0 third component}). The deviation of the last few points of $\dot{W}^{(2)}$ should be attributed to the fluctuation due to the small system size.  }
	\label{Fig: decomposition ratio square lattice}
\end{figure}

The partitioning ratios $\dot{W}^{(i)}/\dot{W}$ ($i=1,2,3$) can also be numerically determined for comparison with the analytic results in Eq.~\ref{Eq: W0 first component}, \ref{Eq: W0 second component} and \ref{Eq: W0 third component}. The result is presented in Fig.~\ref{Fig: decomposition ratio square lattice}, which shows that the ratios converge to their predicted value (dashed lines) as the system is coarse-grained.  

\bibliography{scaling_citations}